\documentclass[preprints,article,submit,moreauthors,pdftex]{Definitions/mdpi}
\usepackage{tabularx}
\usepackage{subfigure}
\usepackage[subfigure]{tocloft}
\usepackage{longtable}
\usepackage{colortbl}
\usepackage{caption}
\usepackage{mdframed}
\usepackage{tcolorbox}
\tcbuselibrary{theorems}
\definecolor{sinbad}{rgb}{0.604,0.8,0.8}
\definecolor{mystic}{rgb}{0.875,0.933,0.933}
\newtcbtheorem[]{mytheo}{Terminology}%
{fonttitle=\color{black}\bfseries,colback=mystic,colframe=sinbad}{th}
\usepackage{multicol}
\usepackage{multirow}

\firstpage{1} 
\pubvolume{1}
\issuenum{1}
\articlenumber{0}
\pubyear{2022}
\copyrightyear{2022}
\datereceived{} 
\dateaccepted{} 
\datepublished{} 
\hreflink{https://doi.org/} 

\Title{When Infodemic Meets Epidemic: a Systematic Literature Review}

\TitleCitation{Title}

\Author{Chaimae Asaad $^{1,2}$*\orcidA{}, Imane Khaouja $^{1}$, Mounir Ghogho $^{1,3}$ and Karim Ba\"ina $^{2}$}

\AuthorNames{Chaimae Asaad, Imane Khaouja, Mounir Ghogho and Karim Ba\"ina}

\AuthorCitation{Asaad, C.; Khaouja, I.; Ghogho, M.;Ba\"ina, K}

\address{%
$^{1}$ \quad TicLab, College of Engineering and Architecture, International University of Rabat\\
$^{2}$ \quad Alqualsadi, Rabat IT Center, ENSIAS, Mohammed V University in Rabat\\
$^{3}$ \quad Faculty of Engineering, University of Leeds, United Kingdom}

\corres{Correspondence: chaimae.asaad@uir.ac.ma}

\abstract{Epidemics and outbreaks present arduous challenges requiring both individual and communal efforts. Social media offer significant amounts of data that can be leveraged for bio-surveillance. They also provide a platform to quickly and efficiently reach a sizeable percentage of the population, hence their potential impact on various aspects of epidemic mitigation. 
The general objective of this systematic literature review is to provide a methodical overview of the integration of social media in different epidemic-related contexts. Three research questions were conceptualized for this review, resulting in over 10000 publications collected in the first PRISMA stage, 129 of which were selected for inclusion. A thematic method-oriented synthesis was undertaken and identified 5 main themes related to social media enabled epidemic surveillance, misinformation management, and mental health.
Findings uncover a need for more robust applications of the lessons learned from epidemic post-mortem documentation. A vast gap exists between retrospective analysis of epidemic management and result integration in prospective studies. Harnessing the full potential of social media in epidemic related tasks requires streamlining the results of epidemic forecasting, public opinion understanding and misinformation propagation, all while keeping abreast of potential mental health implications. Pro-active prevention has thus become vital for epidemic curtailment and containment.
}


\begin{document}


\section{Introduction}

The proliferation of social media content has been the staple of the last decade. Social media platforms have evolved to fulfill numerous and diverse roles, rendering them essential, ubiquitous and a catalyst for change, for better and for worse. 
Social media has been a solemn companion through major crises and events of the past decades, offering a a tool for connection, a space to grieve, and an instrument of outrage. Through epidemics, wars, hurricanes, earthquakes, terrorist attacks and major elections, social media bred everything from support to contention to conspiracy theories. 

Perhaps the most notable event of the decade will be the Covid-19 pandemic. From lockdowns to mask mandates and vaccinations, the world is still grappling with the ongoing management and socio-economic aftermath of this crisis. As the global community braves the third year of the global crisis, with multiple emerging variants, more than 6 million deaths and 517 million reported cases, it is evident that our existence on social media has evidently become an undeniable and fundamental facet of our shared human experience.

Bio-surveillance is defined as "the discipline in which diverse data streams are characterized in real or near-real time to provide early warning and situational awareness of events affecting human, plant,and animal health" including human disease outbreaks \cite{walters2008data}. Efforts directed at facilitating both the early detection and forecasting of disease outbreaks have been increasing in the past two decades. Through the analysis of a variety of data sources, "syndromic surveillance" aims to discern individual and population health indicators before confirmed diagnoses are made \cite{mandl2004implementing}. Infected individuals or populations may exhibit behavioral patterns, symptoms, signs, or laboratory findings that can be tracked \cite{mandl2004implementing}.
Social media platforms offer significant amounts of data that can be used in both bio-surveillance and syndromic surveillance of epidemics and outbreaks. Understanding how social media shapes our experiences in times of crises, and characterizing the roles social media fulfill during epidemics would allow for an improved apprehension of how to efficiently utilize such a crucial tool, and may ultimately hold the key to curb the death toll and prevent devastating consequences.

During epidemics and outbreaks, mistrust of governments and health workers, misinformation, and rumors present challenges to containment and can have a negative impact on mitigation efforts. The particular vulnerability caused by the fear and uncertainty surrounding epidemics, especially amid experiences of loss, renders many social media users highly suggestible and at risk for fake news acceptance and dissemination. The significant financial and medical burden imposed by outbreaks and epidemics, in addition to the substantial challenges arising in their progression and aftermath further complicates the mental health toll they take on the population affected and on vulnerable communities.
Although many literature reviews have shown interest in the subject of the roles social media fulfill during times of crisis in the last decades, a gap exists for this systematic review's research questions. This review's aim is to examine the aspects of the 'epidemic-social media' relationship and categorize its various aspects, as well as analyze if and how it can contribute to an improved management of epidemics. In light of the current state of public health worldwide, it is vital to understand how a tool as influential as social media can shape the population's response in time of crisis and how it can be harnessed to mitigate risks. 

This paper is a systematic literature review aiming to study the literature's take on the relationship between epidemics and social media's impact. This relationship is outlined based on three research questions highlighting (i) the management aspects of epidemics, (ii) the proliferation of misinformation, and (iii) the potential impact on mental health: 
\begin{itemize}
    \item \textbf{RQ1.} Can social media be harnessed for epidemic management and mitigation?
    \item \textbf{RQ2.} Can social media be used for misinformation management during epidemics?
    \item \textbf{RQ3.} Can social media be integrated in aspects of public mental health management during epidemics?
\end{itemize}

In order to accurately frame the concepts discussed in this systematic literature review, we follow the terminological distinctions outlined below by the Dictionary of Epidemiology \cite{porta2008dictionary} :
\begin{mytheo*}{}
\textit{\textbf{Epidemic :}} \textit{"The occurrence in a community or region of cases of an illness, specific health-related behavior, or other health-related events clearly in excess of normal expectancy."}

\vspace{5pt}
\textit{\textbf{Pandemic :}} \textit{"An epidemic occurring worldwide or over a a very wide area, crossing international boundaries, and usually affecting a large number of people."}

\vspace{5pt}
\textit{\textbf{Outbreak :}} \textit{"An epidemic limited to localized increase in the incidence of a disease."}

\vspace{5pt}
\textit{\textbf{Endemic disease :}} \textit{"The constant presence or usual prevalence of a disease or infectious agent within a given geographic area or population group."}

\vspace{5pt}
\textit{\textbf{Infectious disease :}} \textit{"An illness due to a specific infectious agent or its toxic products that arises through transmission of that agent or its products from an infected host."}
\end{mytheo*}

The contributions of this review are manifold:

\begin{itemize}
    \item A systematic categorization and summary of existing methods of epidemic surveillance and forecasting. 
    \item A systematic categorization and summary of existing methods of understanding public opinion and information dissemination on social media during epidemics.
    \item A systematic categorization and summary of existing methods of misinformation detection and characterization on social media during epidemics.
    \item A systematic analysis of the impact of social media on mental health during epidemics.
    \item A systematic analysis of findings for the identification of potential research directions for an improved leveraging of social media for epidemic management and mitigation. 
    \item A systematic analysis of findings for the identification of potential research directions for effective curtailment of fake news propagation and negative impact during epidemics.
    \item A systematic analysis of findings for the identification of potential research directions for effective curtailment of social media's impact on mental health during epidemics.
\end{itemize}

The remainder of this paper is organized as follows. Related surveys are detailed in \textbf{\textit{Section \ref{relatedwork}}} to explain the research gap and need for this systematic review. Methods pertaining to the search strategy and extraction process are detailed in \textbf{\textit{Section \ref{methods}}}.
Based on the comprehensive and systematic survey and investigation of existing methods used to answer the research questions, an overall picture on the current research frontiers is outlined in the form of results and syntheses presented and analyzed in \textbf{\textit{Section \ref{results}}}. Discussion of the major issues and practical implications as well as identified directions for future research are presented in \textbf{\textit{Section \ref{discussion}}}. Final conclusions are summarized in \textbf{\textit{Section \ref{conclusion}}}.

%
\section{Related Surveys}\label{relatedwork}

The literature has taken a special interest in social media's role in times of crises, resulting in several works studying the scientific contributions made to this subject. 

The authors of \cite{charles2015using} conducted a systematic literature review (SLR) aiming to examine the potential of using social media to support and improve public health. This SLR studied two research questions: \textit{(i) Can social media be integrated into disease surveillance practice and outbreak management to support and improve public health?} and \textit{(ii) Can social media be used to effectively target populations, specifically vulnerable populations, to test an intervention and interact with a community to improve health outcomes?}

The scope of this review included outbreaks resulting from both infectious and non-infectious diseases, and covered works published in Pubmed, Embase, Scopus and Ichushi-web ranging from 2008 to 2013 (\textbf{Tab. \ref{surveys}}). The papers reviewed by this SLR included various social media sites as well as discussion forums and blogs. The main findings of this SLR highlight:

\begin{itemize}
    \item The particularly challenging nature of translating research using social media for bio-surveillance into practice. 
    \item The lack of an ethical framework for the integration of social media into public health surveillance systems.
    \item The retrospective nature of many studies on infectious diseases potentially highlighting the ease in post-outbreak prediction in comparison with prospective studies. 
    \item The under-representation of social media analytics in active surveillance.
    \item Knowledge of the population's characteristics and way of using social media is a critical part of successful intervention and surveillance.
    \item The impact of the potential lack of population representativeness in the use of social media to detect and track disease outbreaks has not been adequately researched.
\end{itemize}

The authors noted amongst their concluding remarks the effectiveness of social media in supporting and improving public health and in identifying target populations for intervention. They also recommended identifying opportunities that enable public health professionals to integrate social media analytics into disease surveillance and outbreak management practices.

\renewcommand{\arraystretch}{2.25}
\begin{small}
\begin{longtable}{p{3cm}p{4.5cm}p{3cm}p{2cm}}
\caption{\textbf{Search Characteristics of Related Surveys}}\\
\label{surveys}\\
\rowcolor[rgb]{0.604,0.8,0.8}\textbf{ Review Paper } & \textbf{ Databases Searched } & \textbf{ Time Range } & \textbf{ Total papers reviewed }\\
\rowcolor{white} ~ & ~ & ~ & ~ \\[-5ex]
\rowcolor[rgb]{0.961,0.961,0.961}
Charles-Smith et al. (2015) \cite{charles2015using}  & PubMed, Embase, Scopus and
Ichushi Web & January 2008 – February 2013 & Total = 1514 \newline Included = 60 \\
\rowcolor{white}
~ & ~ & ~ & ~ \\[-5ex]
\rowcolor[rgb]{0.851,0.922,0.922}
Eismann et al. (2016) \cite{eismann2016collective}  & ACM Digital Library, AIS Electronic Library, EBSCOhost, IEEE Xplore Digital Library, JSTOR, ScienceDirect and the Social Science Citation Index & - 29 October 2015 & Total = 3,746 \newline Included = 68 \\
\rowcolor{white}
~ & ~ & ~ & ~ \\[-5ex]
\rowcolor[rgb]{0.961,0.961,0.961}
Tang et al. (2018) \cite{tang2018social}  & PubMed/MEDLINE, PsycINFO, CINAHL Plus, ProQuest and EBSCOhost & January 1, 2010 - March 1,
2016 & Total = 569 \newline Included = 30 \\
\rowcolor{white}
~ & ~ & ~ & ~ \\[-5ex]
\rowcolor[rgb]{0.851,0.922,0.922} 
Abdelhamid et al. (2021) \cite{abdulhamid2021survey}  & Wiley-Blackwell Full Collection, Elsevier/Science Direct, Business Source Premier and Palgrave McMillan databases & 2002 - 2016  & Total = 67 \newline Included = 49 \\
\end{longtable}
\end{small}

The authors present in \cite{eismann2016collective} a systematic literature review focusing on the research question: \textit{What disaster-related collective behavioural phenomena have been observed in social media so far?"} They employ the United Nations Office for Disaster Risk Reduction's definition of a disaster: "serious disruption(s) of a community or society involving widespread human, material, economic or environmental losses and impacts which exceed $[$...$]$ the ability of the affected community or society to cope with using its own resources" \cite{undrr2009}. 

The data used in this SLR originates from several sources such as ACM Digital Library and IEEE Xplore and covers papers published until the date of the last search which occurred on the 29th October of 2015 (\textbf{Tab. \ref{surveys}}) .

The purpose of this SLR involves gaining a better view on phenomena known as "collective behaviour" in the particular case of disasters, given that they emerge when individual actions are embedded into a social context through social media. 
The main findings of this SLR are: 
\begin{itemize}
    \item Sharing and obtaining factual information is the primary function of social media usage consistently across all disaster types, but secondary functions vary.
    \item Disaster management activities are not restricted to individual phases of the disaster management lifecycle in social media. 
    \item The duration, scope, and magnitude of disasters influence the extent of social media usage in a disaster, but not necessarily the structure and function of usage. 
    \item Different actor types make use of social media in similar ways, but perceive different conditions and restrictions for social media usage in disaster situations.
    \item Social media enable members of the population to reach formerly inaccessible actors, but do not ensure two-way communication.
    \item Social media integrate unspecified and wider audiences into disaster communication, which can lead to group emergence.
    \item The features of social media platforms determine the structure and function of collective behavior on these platforms in disasters.
\end{itemize}

In the systematic literature review presented in \cite{tang2018social}, the authors examine the role social media plays in relaying information during emerging infectious diseases (EIDs) outbreaks and identifies the major approaches and assesses the rigors in published research articles on EIDs and social media. This SLR used multiple literature databases such as Pubmed and PsycInfo and covered a timeframe ranging from 2010 to 2016 (\textbf{Tab.\ref{surveys}}).

Amongst the main findings of this SLR is the identification of three major approaches from the reviewed literature:
\begin{itemize}
    \item Assessment of the public's interest in and responses to EIDs.
    \item Examination of organizations' use of social media in communicating EIDs.
    \item Evaluation of the accuracy of EID-related medical information on social media.
\end{itemize}

Additionally, the authors discuss in this SLR the challenges they believe as dominating the field: a lack of theorization and a need for more methodological rigor.

The term emerging infectious disease (EID) refers to both new infectious diseases appearing in the last 20 years or re-emerging infections \cite{oaks1992emerging}. Examples include newly identified species of pathogens such as Severe Acute Respiratory Syndrome (SARS), pathogens affecting new populations such as West Nile Virus, or reemerging outbreaks of measles and drug-resistant tuberculosis. 

The literature review presented in \cite{abdulhamid2021survey} studies existing research evidence on the use of social media use in emergency management. 

The authors of this review use the interpretivist approach as a method of qualitative inquiry and conducted their literature review following the recommendations and methods of literature search and review specified in \cite{vom2015standing}. They also combine the review approaches followed by a literature review paper on social media applications and management (\cite{luna2018social}) and a retrospective review on social media in emergencies and its research with a special emphasis on use patterns, role patterns and perception patterns (\cite{reuter2018fifteen}). The authors base this choice of methodology on the belief that "research blossoms when scholars reuse the methods and knowledge developed by peers". 

The data sources used in this review include Wiley-Blackwell Full Collection and Elsevier, and include scientific contributions published between 2002 and 2016.
The findings of this literature review highlight clusters representing three main themes around the use of social media in emergencies: Information Sharing, situational awareness for decision making and collaboration among citizens, and emergency management organisations, aid agencies as well as digital volunteers. The authors of this review conclude by noting both the potency of social media as a multidimensional tool for reporting, organising and raising global awareness, as well as the need for further empirical evidence to quantify the extent of such potency. The systematic literature review presented in the remainder of this paper differs from the works aforementioned and aims to fill a different gap in the literature. The inclusion of non-infectious diseases and other health risk behaviors (\cite{charles2015using}), the study of disasters in general and the focus on collective behavior (\cite{eismann2016collective}) , the inclusion of multiple infectious diseases both new and reemerging (\cite{tang2018social}), and the generalized perspective on emergency situations (\cite{abdulhamid2021survey}) are all broader and differing scopes than ours. Major differences extending beyond the research questions of our SLR include the time range of the reviewed literature, the volume of data and the databases searched. 

The focal point of this work emphasizes epidemics and pandemics, specifically those having occurred in the last two decades. The proliferation of social media content in the chosen time frame and the global impact of pandemics allows us to study the role social media plays on a bigger scale and gauge its effectiveness or lack thereof. It also allows us to draw more robust conclusions. 
In light of the considerable impact the Covid-19 pandemic has had on the world, it is evident that pandemic preparation and mitigation protocols need to be adjusted to deal with the special challenges that accompany the technological revolution taking place. It is vital to have effective ways to exploit the full potential of social media without risking the toll it could potentially take on users' mental health.
The systematic literature review presented in this paper covers these important aspects of the relationship between pandemic mitigation and social media as well as the role the infodemic of fake news has played.

\section{Methods}\label{methods}

\begin{figure}
    \centering
    \includegraphics[width=1\textwidth]{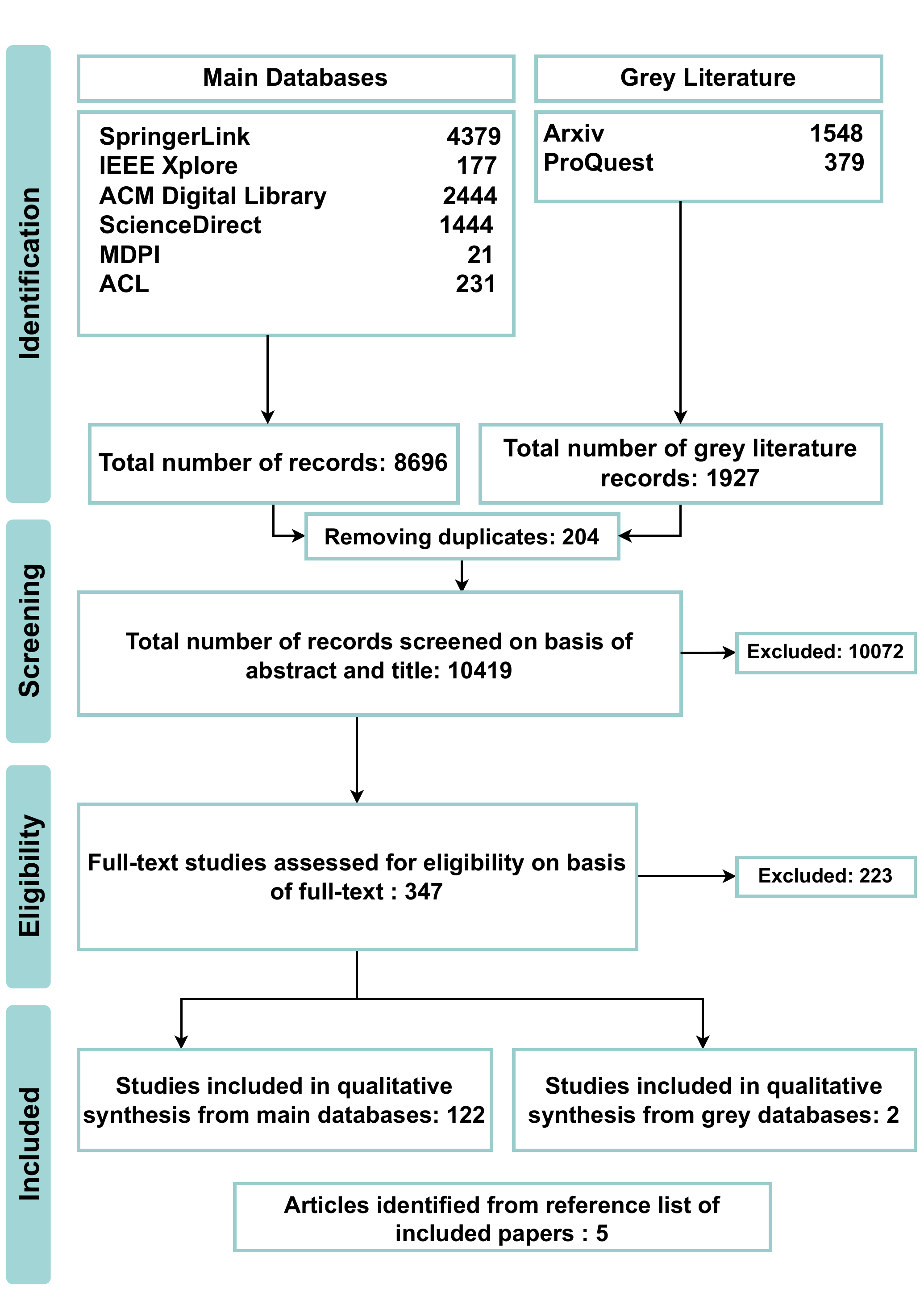}
    \caption{Flow diagram for the selection of the literature reviewed. \textit{The abstract screening process resulted in 347 studies identified for detailed review of full-text articles. After this review, we further excluded studies that did not meet our definition of social media, that discussed epidemic modeling exclusively, or that did not answer the research questions. We identified a total of 124 studies that met our eligibility criteria and addressed the research questions. 5 additional papers identified through the reference lists of selected papers were added. The selection process took almost a year to complete because of the volume of the literature screened.}}
    \label{slr_flow_diagram}
\end{figure}
This systematic review builds upon the preferred reporting items outlined in the PRISMA (Preferred Reporting Items for Systematic Reviews and Meta-Analysis) statement in an effort to properly assess the impact of social media during epidemics \textbf{(Prisma Checklist)}.

The process of conducting a systematic literature review (SLR) entails four main phases (\textbf{\textit{Fig. \ref{slr_flow_diagram}}}): Identification, screening, eligibility and data extraction of included publications. Upon formulating the research questions for the systematic literature review, a literature review relating to the main objectives was conducted to validate the research gap. 

In the identification phase, target databases were selected along with the queries and keywords to be used in the search strategy. Inclusion and exclusion criteria were defined to systematically filter papers in the screening phase based on titles and abstracts, and in the eligibility phase based on full-text reading. 

\subsection{Research Questions}

This systematic literature review aims to examine the different aspects of the integration of social media in epidemic management, and to summarize the methods used for that purpose. 
To that end, the three following research questions were formulated:
\begin{itemize}
    \item \textbf{RQ1: Can social media be harnessed for epidemic management and mitigation?} \\
This research question aims to identify potential uses of social media platforms in the context of epidemic management and/or mitigation. 
    \item \textbf{RQ2: Can social media be used for misinformation management during epidemics?} \\
This research questions aims to examine potential methods used in the context of social media misinformation control as it relates to epidemics.
    \item \textbf{RQ3: Can social media be integrated in aspects of public mental health management during epidemics?} \\
This research question aims to discern potential aspects of the relationship between social media and public mental health during epidemics.
\end{itemize}

\subsection{Search Strategy}	

A systematic literature search was undertaken from June 2021 to March 2022.
A collaborative planning and task allocation process was developed, and updated at each stage of the study. The systematic search was conducted across multiple scientific databases: IEEE Xplore, ACM Digital Library, ScienceDirect, MDPI, ACL and SpringerLink. 
Grey literature sources Arxiv and ProQuest were used to complement the search strategy and reduce publication bias by providing a venue to share studies with null or negative results that might otherwise not be disseminated. 

The research questions were used as a guideline to 'roughly' identify the main search keywords. The search terms used included "social media" and "epidemics", with variations depending on the research question's objectives and the database searched.
For research question 1 (RQ1), the results of the query ("social media" AND "epidemics") was complemented by the results of the query ("social media" AND "epidemics" AND "monitoring" AND "tracking"). The combination of these two queries allowed for a balance of result filtering without over-limiting the output.
The query ("social media" AND "epidemics AND "fake news") was used for research question 2 (RQ2). A combination of the queries ("social media" AND epidemics AND "mental health" AND "support system") and ("social media" AND epidemic AND "mental health" AND addiction) was used for research question 3 (RQ3).
These queries were adapted to each database based on the settings allowed in each database. All searches used the parameters \textit{full-text} or \textit{all metadata} in the queries when the database allowed the specification.  All searches covered the time range 2010 - May, $31^{st}$ 2021.
\textbf{\textit{Tab. \ref{numberofpapers}}} details the number of publications retrieved from each database for each research question.

\renewcommand{\arraystretch}{2.25}
\begin{small}
\begin{longtable}{p{6cm}p{2cm}p{2cm}p{2cm}}
\caption{\textbf{Output of Search Strategy for research questions RQ1, RQ2 and RQ3}}\\
\label{numberofpapers}\\
\rowcolor[rgb]{0.604,0.8,0.8}
\textbf{ Database } & \textbf{RQ1} & \textbf{RQ2} & \textbf{RQ3} \\
\rowcolor{white}
~ & ~ & ~ & ~ \\[-5ex]
\rowcolor[rgb]{0.961,0.961,0.961}
IEEE Xplore  & 152 & 7 & 18\\
\rowcolor{white}
~ & ~ & ~ & ~ \\[-5ex]
\rowcolor[rgb]{0.851,0.922,0.922}
ScienceDirect & 1000 & 272 & 172\\
\rowcolor{white}
~ & ~ & ~ & ~ \\[-5ex]
\rowcolor[rgb]{0.961,0.961,0.961}
SpringerLink & 1892 & 347 & 2140\\
\rowcolor{white}
~ & ~ & ~ & ~ \\[-5ex]
\rowcolor[rgb]{0.851,0.922,0.922} 
ACL & 90 & 20 & 121\\
\rowcolor{white}
~ & ~ & ~ & ~ \\[-5ex]
\rowcolor[rgb]{0.961,0.961,0.961} 
ACM Digital Library & 512 & 699 & 1233 \\
\rowcolor{white}
~ & ~ & ~ & ~ \\[-5ex]
\rowcolor[rgb]{0.851,0.922,0.922} 
MDPI & 20 & 0 & 1 \\
\rowcolor{white}
~ & ~ & ~ & ~ \\[-5ex]
\rowcolor[rgb]{0.961,0.961,0.961}
Arxiv & 1544 & 4 & 0\\
\rowcolor{white}
~ & ~ & ~ & ~ \\[-5ex]
\rowcolor[rgb]{0.851,0.922,0.922} 
ProQuest & 226 & 26 & 127 \\
\end{longtable}
\end{small}

\subsection{Study Selection and Data Extraction Strategy}
The Preferred Reporting Items for Systematic Reviews and Meta-Analyses (PRISMA) statement was followed in the selection process. The point of interest in the inclusion criteria included any published full-text research article on the uses of social media in epidemic management for research question 1, the uses of social media in misinformation management for research question 2, and the aspects of public mental health related to social media, for research question 3. At the initial screening stage, apart from duplicates removal, three authors assessed the titles and abstracts against the criteria of inclusion. Publications included after this screening stage are then retrieved in full-text version, and subsequently screened in the eligibility stage. Three of the authors read the full-text articles independently to ascertain their relevance with regard to the search terms and the research aim. All disagreements on inclusions in the screening and eligibility stages were resolved by consensus. 
In order to organize the screening process, RAYYAN\footnote{\url{https://www.rayyan.ai/}} \cite{ouzzani2016rayyan}, a web app facilitating the collaborative review process and screening process for systematic literature reviews, was used by authors to import all articles initially collected and screen them following a "blind on" setting, where decisions and labels of any collaborator were not visible to others. Publications with inclusion disagreements were then identified after dropping the "blind on" setting, and resolved amongst authors.

The inclusion and exclusion criteria were specified and agreed upon by all authors following the aims and objective of the systematic literature review (\textbf{\textit{Tab. \ref{inclusioncriteria}}} ). In order for a publication to be selected, it needs to address the research question(s), and to be published within the agreed upon time range. The publication is excluded if it is not a journal paper, conference proceedings paper or peer-reviewed workshop paper. Papers are also excluded if full-text versions are not available. Publications relating to the HIV epidemic or the tuberculosis (TB) epidemic were excluded to preserve the homogeneity of the review. TB is a bacterial infection with a high burden of disease, especially in developing countries, while HIV (human immunodeficiency virus) is the virus responsible for AIDS (acquired immune deficiency syndrome). Both TB and HIV/AIDS are classified as ongoing worldwide public health issues by the World Health Organization (WHO) and the CDC (Centers for Disease Control and Prevention). Given the particularities of both TB and HIV/AIDS and the high volume of literature review publications related to them (e.g., \cite{taggart2015social}), the authors agreed to consider both outside the scope of this systematic literature review.

\renewcommand{\arraystretch}{2.25}
\begin{small}
\begin{longtable}{p{6.5cm}p{6.5cm}}
\caption{\textbf{Inclusion/Exclusion criteria for study selection process}}\\
\label{inclusioncriteria}\\
\rowcolor[rgb]{0.604,0.8,0.8}
\textbf{ Inclusion Criteria } & \textbf{Exclusion Criteria} \\
\rowcolor{white}
~ & ~ \\[-5ex]
\rowcolor[rgb]{0.961,0.961,0.961}
Within scope of one of the research questions  &  - \\
\rowcolor{white}
~ & ~ \\[-5ex]
\rowcolor[rgb]{0.851,0.922,0.922}
Within time frame 2010 - May 31$^{st}$, 2021 & - \\
\rowcolor{white}
~ & ~ \\[-5ex]
\rowcolor[rgb]{0.961,0.961,0.961}
Relates to an epidemic or pandemic within the last two decades & Tuberculosis, HIV, non-infectious diseases \\
\rowcolor{white}
~ & ~ \\[-5ex]
\rowcolor[rgb]{0.851,0.922,0.922}
Includes the use of a social media site & Online forums, traditional media \\
\rowcolor{white}
~ & ~ \\[-5ex]
\rowcolor[rgb]{0.961,0.961,0.961}
Is a journal, conference or workshop paper & Book, e-book, letter to editor, magazine, abstracts, case reports, comments, reviews, posters \\
\end{longtable}
\end{small}
In the data extraction stage, the final list of papers was analyzed to answer the research questions and extract pertinent information.
The final stage of the PRISMA guidelines were considered in this phase. The following data was extracted from selected papers: author(s), publication year, epidemic studied, social media site used, dataset size, study design, identified method and key findings. All the related data were extracted independently by two investigators. When necessary, differences were resolved by discussing, examining, and negotiating with a third investigator.

\subsection{Quality Assessment}
The quality of included studies in this review was appraised using a set checklist of quality criteria. 
Papers that did not fulfill at least 4 out of the 5 quality criteria are excluded. 
The checklist was defined as follows:
\begin{itemize}
    \item (1) : Are the study objectives clearly defined?
    \item (2) : Are the methods clearly defined and applied?
    \item (3) : Are the methods applied successfully and correctly?
    \item (4) : Are accuracy values and efficiency/confidence levels reported?
    \item (5) : Are limitations clearly reported and adequately represented?
    \item (6) : Do the contributions outweigh the limitations of the study?
\end{itemize}
The quality criteria were formulated based on our understanding of the current state of research in this field and the research gap this systematic review is attempting to fill. The papers were assessed for their ability to answer the research questions and enrich the literature while fulfilling quality standards.

Bias was evaluated in this systematic literature review from 2 aspects. Firstly, risk of bias based on inclusion bias was limited through the use of multiple reviewers. Secondly, publication bias was limited by including grey literature which reports negative and null results. 
In order to enhance the quality of this review, the authors monitored the planned review tasks and ensured continuous progress monitoring. Collaborative worksheets were created to keep track of scheduled tasks and deadlines, and to note pertinent observations. Validation of the extracted data from selected papers was conducted by the authors, and peer-reviewing was maintained at every stage of the systematic review process.

\section{Results}\label{results}
\subsection{Characteristics of Selected Papers}

The search process resulted in a total of \textit{10623} articles distributed over both the main and grey databases used (\textbf{\textit{Fig. \ref{slr_flow_diagram}}}). After the removal of duplicates, \textit{10419} titles remained. Of these, \textit{10072} studies were excluded after the title and abstract screening, as they did not fulfill the criteria.  Of the \textit{347} studies that were full-text screened, \textit{223} did not meet the inclusion/exclusion criteria. A total of \textit{124} studies, as well as 5 additional papers identified from reference lists of included papers, were selected for inclusion in the current review as summarized in \textbf{\textit{Sec. \ref{answers}}}.

\begin{figure}[ht!]
    \subfigure[Total number of articles per year]{\includegraphics[width=0.49\textwidth]{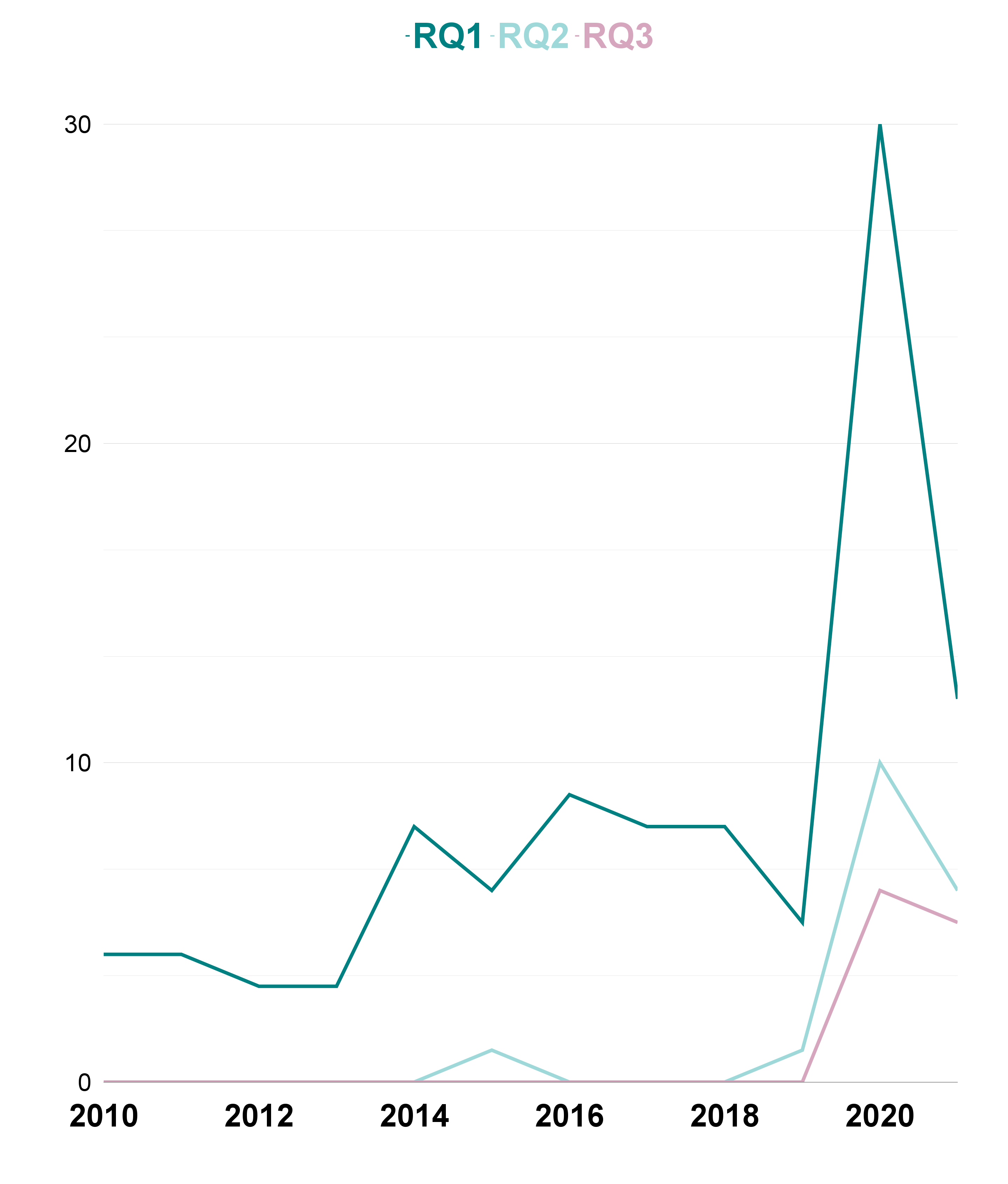}}
    \subfigure[Distribution of papers per type of publication]{\includegraphics[width=0.49\textwidth]{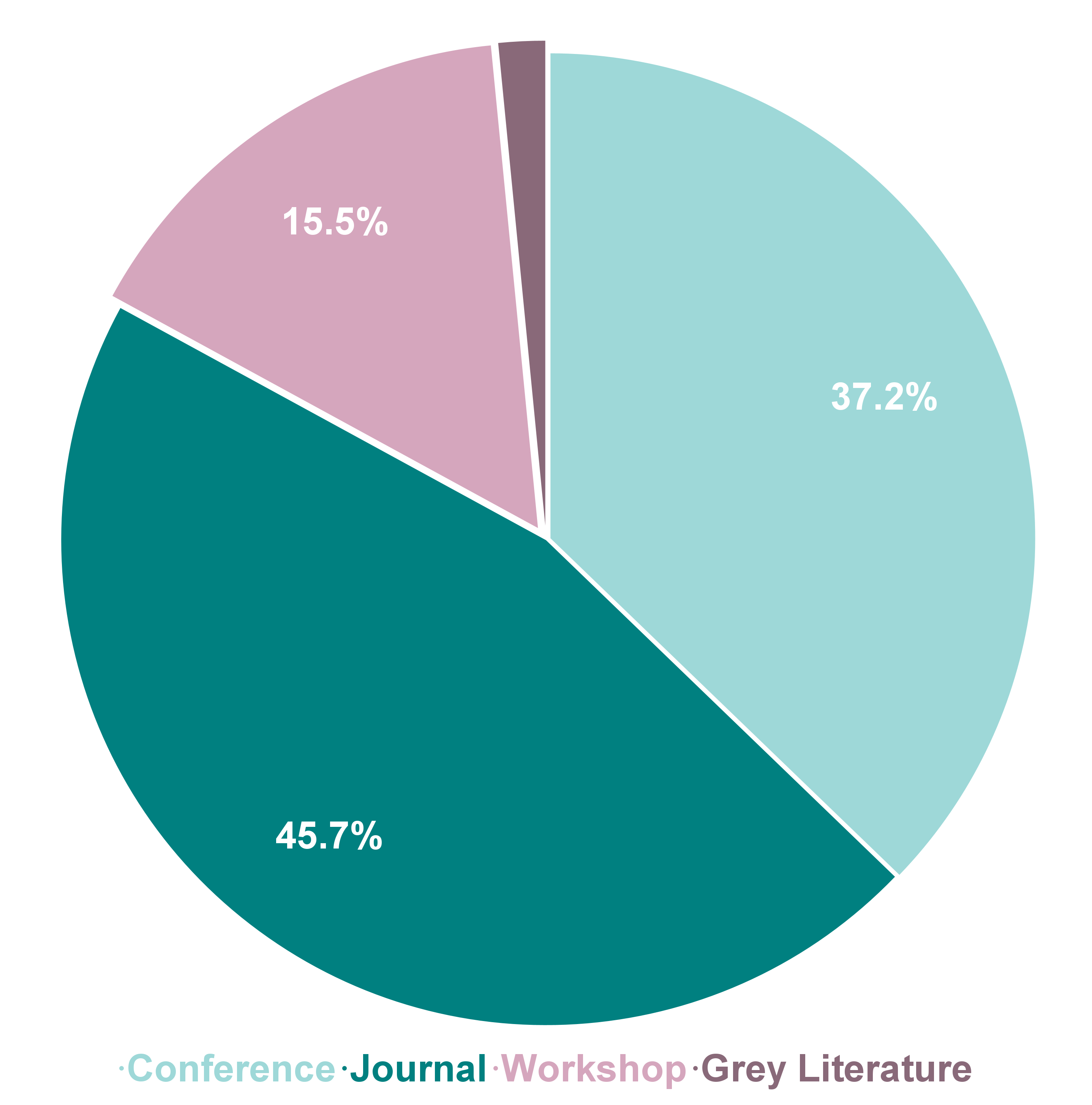}}
    \caption{Statistics on Selected Papers of Systematic Literature Review}
    \label{stats}
\end{figure}

The papers included in the systematic literature review were distributed as follows: 45.7\% are journal papers, 37.2\% are publications of conference proceedings, 15.5\% are workshop publications, while 1.6\% are grey literature (\textbf{Fig. \ref{stats}(b)}).

The publications included span the 2010-2021 period. As can be seen in \textbf{Fig. \ref{stats}(a)}, the number of publications peaks in 2020 for all research questions. This is due to the rapid increase of the Covid-19 related publication rate during the first year of the pandemic. The frequency of publication and the volume of the academic output contributed to the creation of the Covid-19 Open Research Dataset (CORD-19) \cite{wang2020cord}.  

Selected publications answering research question 3 (RQ3) all span the 2020-2021 period. Given the mental health aspect of this particular research question, a potential inference can be made suggesting a very recent interest in mental health as it relates to social media and to epidemics. A similar trend can be seen in research question 2 (RQ2), where selected papers are from 2015, 2019, 2020 and 2021. This can be explained by the emergence of the "fake news" phenomena on social media and its particular increase in times of crisis. Research question 1 (RQ1), which studies the aspects of epidemic management and mitigation using social media, includes the highest number of papers, and spans the entire decade, illustrating continuous and ongoing efforts by the scientific community to harness social media's potential for improved containment measures during epidemics. 
\subsubsection{Epidemics Studied}
The selected literature discussed multiple epidemics \textbf{\textit{Fig. \ref{timeline}}}. In what follows, we present them following a symptomatically indicative categorization.
\begin{itemize}
    \item \textbf{Viral hemorrhagic fevers} (VHFs): \textbf{Dengue Fever} and \textbf{Zika Fever} are mosquito-borne disease caused by the \textit{Dengue Virus} and \textit{Zika Virus}, respectively. Both Dengue and Zika are spread by several species of female mosquitoes of the \textit{Aedes} genus\footnote{A taxonomic rank used in the biological classification of living organisms and viruses. In the hierarchy of biological classification, genus comes above species and below family (As Defined by the International Committee on Taxonomy of Viruses)}.

The number of \textbf{Dengue Fever} cases reported to the World Health Organization (WHO) increased over 8 fold over the last two decades, rising from 505,430 cases in 2000, to over 5.2 million in 2019 \cite{thewho}. The disease is now endemic in more than 100 countries in the regions of Africa, the Americas, the Eastern Mediterranean, South-East Asia and the Western Pacific, with Asia representing ~70\% of the global burden of disease \cite{thewho}. In addition to the escalating risk of major outbreaks, extension to other geographic areas remains a risk \cite{guzman2015dengue}. The threat of a possible outbreak of dengue now exists in Europe, with cases observed on an annual basis in several European countries. The largest number of dengue cases ever reported globally was in 2019. Dengue continues to affect many regions in 2021 \cite{thewho}. 

An outbreak of the \textbf{Zika} virus, first discovered in Uganda in the 1940s, occurred in Brazil in early 2015. In February 2016, the WHO declared the outbreak a \textit{Public Health Emergency of International Concern (PHEIC)}, and by the middle of the year more than 60 countries report cases of the virus, including the United States. Thousands of women infected with the virus while pregnant give birth to babies with congenital conditions. The WHO declared the end of the epidemic in November 2016, but reports of cases were registered well into 2017 \cite{thewho}. To date, a total of 86 countries and territories have reported evidence of mosquito-transmitted Zika infection \cite{thewho}. 
\begin{figure}[H]
    \centering
    \includegraphics[width=1\textwidth]{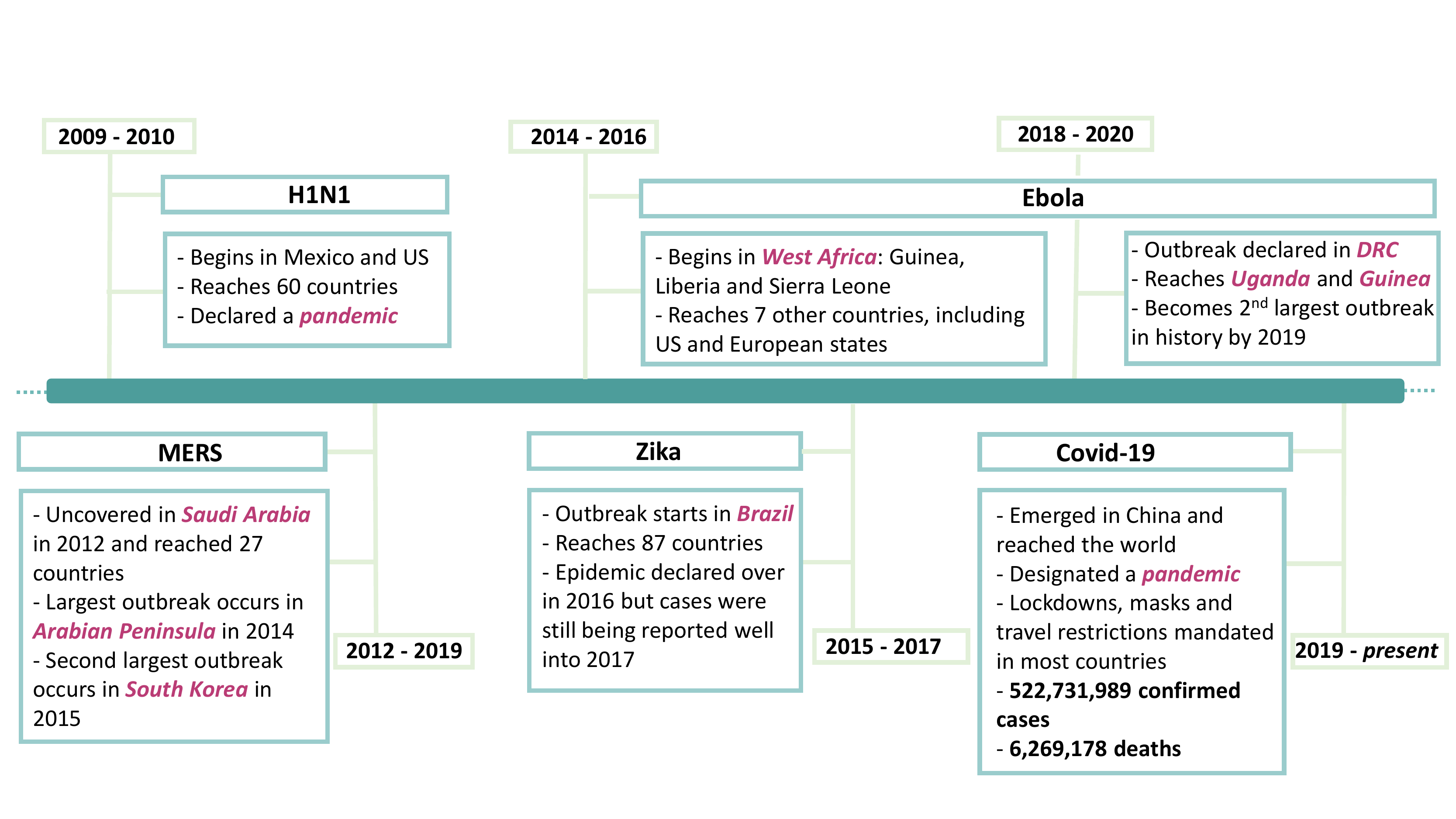}
    \caption{Timeline of the epidemics and pandemics spanning the last decade and included in the Systematic Literature Review. \textit{SARS is pre-2009 and Dengue Fever has caused multiple outbreaks. Both are not illustrated in the timeline, but are included in the SLR.}}
    \label{timeline}
\end{figure}
In early 2014, cases of the \textbf{Ebola} virus were detected in Guinea and soon after in Liberia and Sierra Leone. This instance marked the first time the disease moved into densely populated urban areas, allowing for rapid transmission. The outbreak, declared a PHEIC by the WHO in August 2014, eventually spread to seven other countries, including several European states and the United States, causing more than eleven thousand deaths in total. The 2014–2016 outbreak in West Africa was the largest and most complex Ebola outbreak since the virus was first discovered in 1976 \cite{thewho}. There were more cases and deaths in this outbreak than all others combined. The hardest-hit countries declare themselves Ebola-free in June 2016.
In August 2018, the Democratic Republic of Congo (DRC) declared an outbreak of the Ebola virus in the country’s northeast. Several cases are later reported across the border in Uganda. By June 2019, the Ebola outbreak became the second largest in history, and in July the WHO declared it a PHEIC. More than 3,400 people were infected with the virus, and close to 2,300 died. In June 2020, the WHO declared the end of the outbreak, however, new cases emerged afterward in the DRC and in Guinea \cite{thewho}.
    \item \textbf{Influenza-like Illness} (ILI) is a non-specific respiratory illness characterized by fever, fatigue, cough, and other symptoms. Cases of ILI can be caused either by influenza strains or by other viruses such as Coronaviruses. 
    \begin{itemize}
        \item \textbf{Influenza:} is an acute respiratory infection caused by influenza viruses. It represents a global and year-round disease burden, and causes illnesses that range in severity and sometimes lead to hospitalization and death. Seasonal influenza epidemics mainly caused by influenza A and B viruses result in 3 to 5 million cases of severe illness and a death tool starting at 300,000 deaths worldwide each year \cite{thewho}. In temperate climates, seasonal epidemics occur mainly during winter, while in tropical regions, influenza may occur throughout the year, causing outbreaks more irregularly \cite{bloom2013latitudinal}.
The influenza A virus subtype strain \textbf{H1N1}, labeled and commonly referred to as the \textit{swine flu}, began to spread in early 2009 in Mexico and the United States. Unlike other strains of influenza, H1N1 disproportionately affected children and younger people. The WHO declared it a PHEIC in April 2009, then designated the spread of H1N1 a pandemic in June after the virus reached more than 74 countries and territories. It is estimated that the global death toll of H1N1 ranged from 151,700 to 575,400. The WHO announced the pandemic’s end in August 2010, though the virus continues to circulate seasonally \cite{thewho}.
    \item \textbf{Coronaviruses}: The \textbf{Severe Acute Respiratory Syndrome} (SARS) coronavirus, part of a family of viruses that commonly cause respiratory symptoms, is first identified in late 2002 in southern China. The 2002–2004 outbreak of SARS caused over 8,000 infections and resulted in at least 774 deaths across 29 countries and territories \cite{thewho}. It was "the first severe and readily transmissible new disease to emerge in the 21st century" and showed a clear capacity to spread along the routes of international air travel \cite{thewho}.
The \textbf{Middle East Respiratory Syndrome} (MERS), is a zoonotic virus transferred to humans from infected dromedary camels. It is contractable through direct or indirect contact with infected animals. The largest outbreak occurred on the Arabian Peninsula in the first half of 2014, with the Saudi city of Jeddah as its epicenter. In 2015, South Korea was home to the second-largest outbreak. In total, 27 countries have reported cases of the viral respiratory disease in the following years. The virus has a relatively high fatality rate leading to 858 known deaths due to the infection and related complications of the roughly 2,500 people diagnosed since its discovery \cite{thewho}. 
A novel \textbf{Coronavirus disease} (COVID-19) caused by the SARS-CoV-2 virus emerged in China’s Hubei Province in late 2019, spreading rapidly to other parts of China. The virus rapidly spread throughout the rest of the world, and in March 2020, the WHO designated the outbreak a PHEIC and a pandemic. After more than two years, the official death toll from the COVID-19 disease surpasses six million people, though the actual number is believed to be much higher.  The virus spreads through respiratory droplets and aerosols \cite{thewho}.
    \end{itemize}
\end{itemize}




\begin{figure}
    \centering
    \includegraphics[width=1\textwidth]{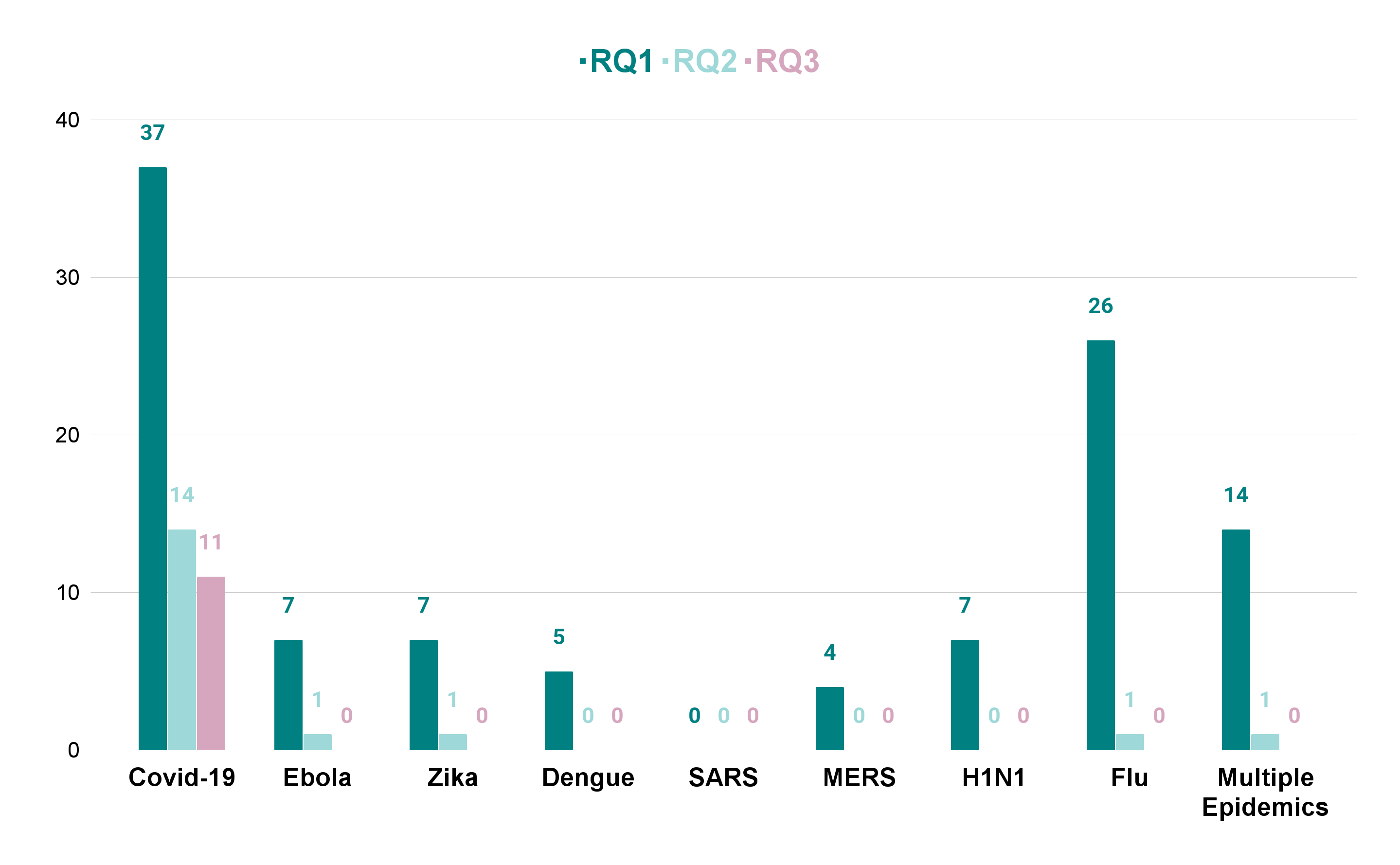}
    \caption{Number of selected publications pertaining to each epidemic included in the systematic literature review.}
    \label{epidemics_rq}
\end{figure}

The highest number of selected publications for all research questions relates to Covid-19, followed by the flu (\textbf{Fig. \ref{epidemics_rq}}). This trend is due, in part, to the volume of the Covid-19 research output, which was motivated by the urgent and challenging nature of the situation.

\subsubsection{Social Media Platforms Used}

Several social media platforms were used in the literature selected for this systematic review. Twitter is one of the most widely used platforms for sharing "micro-blogs". These short messages are called tweets, and can take up to 280 characters. Weibo, on the other hand, is a popular platform to share and discuss individual information and life activities, as well as celebrity news, in China.

\begin{figure}
    \centering
    \includegraphics[width=1\textwidth]{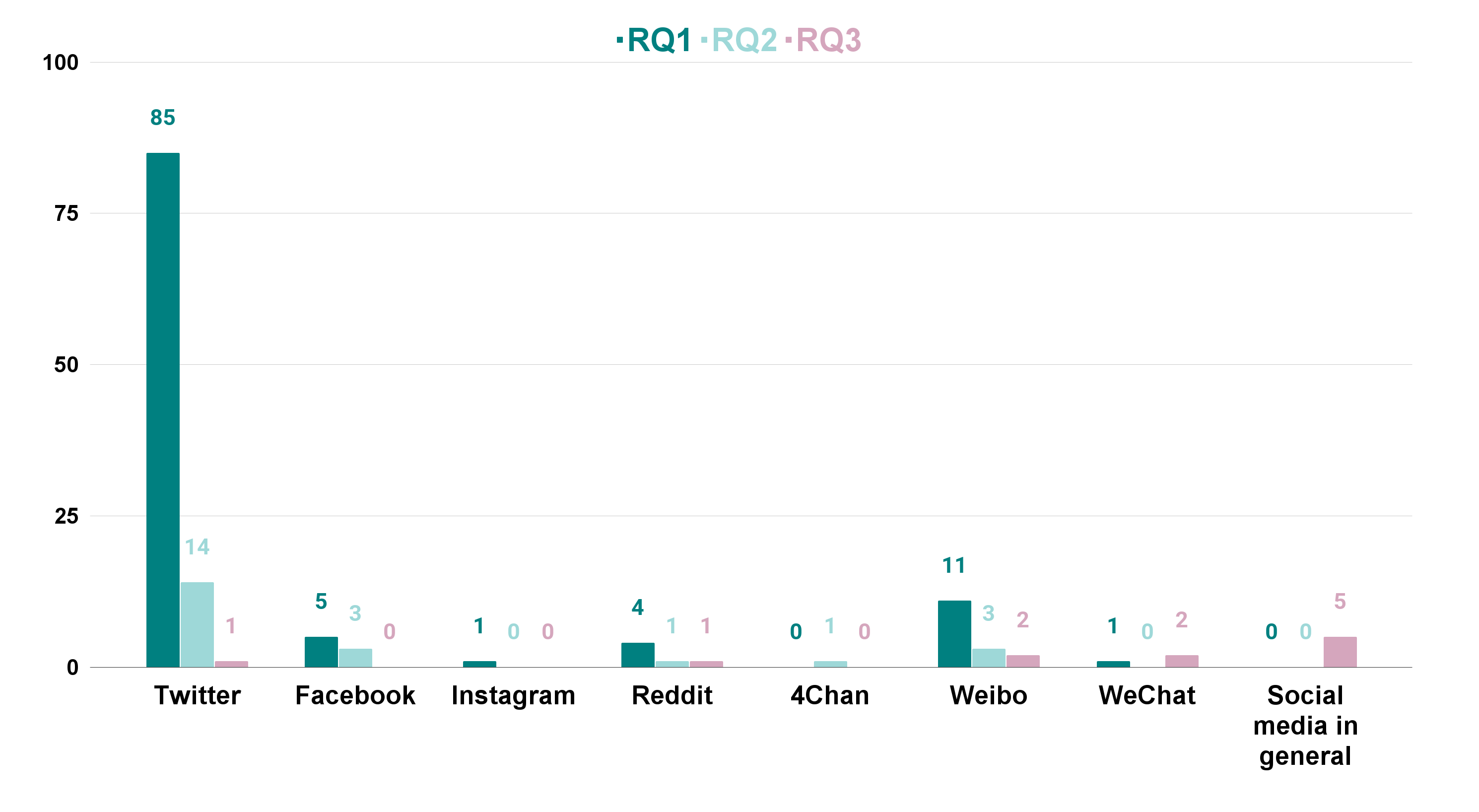}
    \caption{Number of selected publications using each social media platform included in the systematic literature review.}
    \label{socialmedia_rq}
\end{figure}

As can be seen in \textbf{Fig. \ref{socialmedia_rq}}, Twitter, followed by Weibo, seem to be the platform of choice for most works aiming to study epidemic monitoring and mitigation through social media (RQ1), and epidemic related misinformation on social media (RQ2). For epidemic and social media related mental health aspects, most works seem to take a generalist approach rather than a platform-specific one. 

Compared with other social media sites such as Facebook and Instagram, which predominantly include heterogeneous posts, Twitter offers a more concise "micro-blog" format particularly "easy" to analyze. This particular feature may potentially attract works aiming to conduct linguistic analysis or classification tasks.
These differences can also be interpreted geographically, as Twitter is more popular in the United States, while Europe is the largest continent on Facebook with 232 million users, and Weibo is most frequently used in China. Availability of APIs to crawl data is also a major factor in choosing specific social media platforms as data sources.

\subsection{Answers to Research Questions}\label{answers}
    A thematic analysis of the literature included was conducted with the aim of identifying the main themes of each research question. This chosen synthesis method allows for a clearer organization of the literature. The themes are identified following the objectives of the paper and its results. For each theme, papers are organized by method, social media platform used and epidemic studied. Methods are grouped into parent groups. For instance, content analysis includes automated, linguistic, thematic, qualitative or quantitative analysis, while dictionary-based classification entails a lexicon-based classification. Machine learning classification includes conventional machine learning models, while deep learning entails methods based on artificial neural networks with representation learning. 
\subsubsection{RQ1: Can social media be harnessed for epidemic management and mitigation?}
Social media platforms offer significant amounts of data, which can potentially be highly useful for bio-surveillance and syndromic surveillance of epidemics and outbreaks.  

Two main themes were identified in the selected papers addressing how social media can be used in epidemic management, namely: (a) epidemic surveillance and forecasting, and (b) public opinion understanding. Methods used in each of these themes were classified into groups depending on the aim and results of the study.

\paragraph{\textbf{a. Epidemic surveillance and forecasting}}
An important body of research tackled epidemic surveillance and forecasting through social media.

Using dictionary based classification, \cite{kwon_defining_2020} aimed to define and quantify the prevalence and evolution of facets of social distancing during the COVID-19 pandemic in the US in a spatio-temporal context. As a result, tweets containing keywords were grouped into six social distancing facets such as implementation, purpose, social disruption and adaptation. \cite{rao_retweets_2020} used official twitter accounts (CDC, WHO, NIH) to collect and then analyze their communications through tweets. These tweets were classified as "reassuring" or "alarming" based on contextually similar keywords provided by word2vec, an embedding technique based on shallow neural networks. A dengue active surveillance system framework was proposed in \cite{othman_proposed_2016} as a prediction system aiming to improve upon the passive surveillance systems available in Malaysia. This work used data collected from both weather and flood information and social media, and aggregated them in order to be processed and filtered using keywords. The use of machine learning models for the prediction of Dengue cases was also proposed. An application of a spatio-temporal active surveillance system for Dengue is presented in \cite{gomide_dengue_2011}  using four dimensions associated with Twitter data: volume, location, time and public perception. \cite{yom-tov_ebola_2015} collected data from several sources including Twitter from both the main Ebola infected countries in Africa and from across the world between 2013 and 2014. Based on their correlation analysis, the authors concluded that the number of Twitter messages related to Ebola was more correlated with the number of news articles than it was with the number of cases of the disease, and that none of the data sources they used could have provided an alert more than a week before the World Health Organization. In \cite{feng_weibo_2016}, an investigation into using keyword-based extracted data from Weibo could be used as a monitoring platform for Ebola in China was conducted.

Filtering and normalization \cite{szomszor__2010} as well as quantitative analysis based on word counts was used on Twitter data for the surveillance of H1N1 \cite{de_quincey_early_2009} and Zika \cite{abouzahra_twitter_2021}. 
Geo-tagged flu-related twitter data streams were filtered and collected for mathematical modeling of flu trends in \cite{chon_modeling_2015}. Time series of Twitter messages reporting a combination of symptoms matching influenza-like-illness was performed in \cite{stilo_predicting_2014}. Keyword analysis was used for flu risk surveillance \cite{hwang_spatiotemporal_2013} and condition aggravation \cite{talvis_real-time_2014}. Regression was also used for tracking and forecasting Influenza \cite{lampos_enhancing_2017,comito_improving_2018,lampos_flu_2010,zhang_social_2015}, and Zika \cite{masri_use_2019}.
\cite{byrd_mining_2016} conducted twitter extraction using keywords and sentiment analysis for Influenza detection and surveillance. Keywords were also used in addition to time series in order to demonstrate the role of social media for Influenza early warning and upcoming spike detection \cite{kostkova__2014}. Segmentation of the series of differences into an epidemic and a non-epidemic phase using a four-stage Markov switching model. Real time early-stage influenza detection with emotion factors was performed using a segmentation of the series of epidemic/non-epidemic differences through a four-stage Markov switching model \cite{sun_real_2014} . 
Manual content analysis \cite{gui_managing_2017} and keyword definition were used for risk assessment for Zika and public health surveillance of multiple epidemics \cite{romano_challenges_2016,ji_epidemic_2012,chaudhary_use_2017}.
Several conventional machine learning models were used for epidemic surveillance and monitoring.
A proposed event monitoring methodology aiming to handle data scarcity and noise using social media data was applied to Dengue-related data in  \cite{souza_evolutionary_2014}. Lazy associative classification was used to classify tweets into different categories (personal experience, ironic/sarcastic, information, opinion and campaign), while a genetic algorithm was used to obtain the coefficients that maximize the correlation between collected data and event-related statistics. Similarly, lazy association was also used to identify hot spots associated with Dengue from social media trajectories \cite{souza_infection_2016}.
\cite{collier_omg_2011} created guidelines for tagging self protective behaviour and applied them to tweets related to influenza like illness. Using two conventional supervised machine learning classifiers (SVM and Naive Bayes), tweets were classified into 4 self-reported protective behaviour categories and a self-reported diagnosis. Results were compared with H1N1 positive tests reported by the CDC in the United States. 

Several papers leveraged different conventional machine learning models to approach social media enabled flu detection (\cite{wakamiya_after_2016,zhang_detecting_2014,culotta_towards_2010}, by tracking flu activity \cite{zadeh_social_2019}, and predicting prevalence of Influenza-Like Illness \cite{zhang_predicting_2017}. Tweet classification was leveraged in many works to distinguish awareness from infection \cite{lamb_separating_2013}, and predict seasonal flu trends \cite{aramaki_twitter_2011,achrekar_online_2012}.
A location-specific Flu state detection model was proposed, in the series of papers \cite{xiao_detecting_2016,sun_real_2014,sun_hybrid_2015}, with personal emotional factors and semantic information in order to identify the most probable city or province in China where a flu outbreak could take place. Flu-related microblogs were extracted in real-time using a Support Vector Machine (SVM) filter. Posts were classified using association rule mining, and transition time of the flu was detected using conditional random field model. 

Twitter data from Ebola and MERS outbreaks was used in \cite{rudra_classifying_2017,rudra_classifying_2018} to automatically classify tweets into different disease related categories such as: treatment, prevention, symptom, transmission, death report, etc. A similar classification method is performed in \cite{ghosh_identifying_2019} into similar informative categories on Ebola, MERS and Dengue. Supervised text classification was used in addition to topic modeling in \cite{sousa_vazadengue_2018}, while a combination of Support Vector Machines, Naïve Bayes and Logistic Regression was used for infectious disease analytics in \cite{hong_social_2018} for multiple epidemics. Naive Bayes was used in \cite{swain_analysis_2017}. Keyword filtering, text classification, sentiment analysis and Latent Dirichlet allocation (LDA) based topic modeling techniques were applied in \cite{jain_effective_2018} to filter relevant topics related to symptomatic manifestation and prevention of mosquito-borne diseases. 

Deep learning techniques were applied in \cite{mundotiya_nlprl_2020,wang_using_2017} for epidemic monitoring. Few shot learning was used in \cite{lwowski_covid-19_2020} to fine-tune a semi-supervised model built on an unlabeled Covid-19 dataset. A BERT-based model for the Personal Health Mention Identification was employed in \cite{amin_detecting_2020} for the detection of disease-infected individuals in tweets, and a spatial analysis is performed for the identification of disease-infected regions. Classification of zika and ebola related tweets using contextualized word2vec was performed in \cite{khatua_tale_2019}. 
Twitter data was used in \cite{huo_modeling_2020} along with a statistical analysis built on an H1N1 model with best fit parameter values identified by gray wolf optimizer and least square method to show the correlation and implications that Twitter reports have on the control of infectious diseases. Similarly, mathematical modeling was used in \cite{huo_modeling_2016} to understand the influence of Twitter on the spread of H1N1. In order to predict the Covid-19 epidemic peaks and sizes, \cite{nie2021dynamical} introduced information entropy into the quantification of the impact of social network information. Using a semi-supervised multilayer perceptron (MLP) to mine epidemic features, and an online stochastic training algorithm, \cite{zhao_online_2020} illustrated Influenza H1N1 disease outbreak forecasting and characterized disease progress at individual-level.

Multiple papers used social network theory and social network analysis for the prediction of infected groups and early detection of contagious outbreaks in social media \cite{spurlock_predicting_2020,haupt2021characterizing,bagavathi_social_2018,martin_leveraging_2011}. Topic modeling techniques such as LDA were used for epidemic intelligence \cite{zheng_uncovering_2020-1,shin2020risk,chen_flu_2014,fu_detect_2014,nolasco_subevents_2019}, to detect major epidemic related events \cite{missier_tracking_2016}, to monitor information spread \cite{odlum_what_2015}, and to rank epidemic related tweets \cite{diaz-aviles_tracking_2012}.


\renewcommand{\arraystretch}{2.25}
\begin{small}
\begin{longtable}{p{4.5cm}p{3cm}p{3cm}p{2cm}}
\caption{\textbf{Summary of methodologies used in studies addressing research question 1: Epidemic Surveillance and Forecasting} \textit{(ML: Machine Learning, DL: Deep Learning, SNA: Social Network Analysis.)}}\\
\label{table:rq1surveillance}\\
\rowcolor[rgb]{0.604,0.8,0.8}\textbf{Method} & \textbf{Epidemic Studied} & \textbf{Social Media Used} & \textbf{References} \\
\rowcolor{white}\multirow{3}{*}~ & ~ & ~ \\[-5ex]
\multirow{3}{*}{\cellcolor[rgb]{0.961,0.961,0.961}}{\textbf{Dictionary-based Classification}} & {\cellcolor[rgb]{0.961,0.961,0.961}} \textbf{Covid-19} & {\cellcolor[rgb]{0.961,0.961,0.961}} \textbf{Twitter} & {\cellcolor[rgb]{0.961,0.961,0.961}} \cite{kwon_defining_2020,rao_retweets_2020} \\ 
\rowcolor{white}\multirow{3}{*}{\cellcolor[rgb]{0.961,0.961,0.961}} & ~ & ~ \\[-5ex]
\multirow{-2}{*}{\cellcolor[rgb]{0.961,0.961,0.961}} & {\cellcolor[rgb]{0.961,0.961,0.961}} \textbf{Dengue Fever} & {\cellcolor[rgb]{0.961,0.961,0.961}} \textbf{Twitter} & {\cellcolor[rgb]{0.961,0.961,0.961}} \cite{othman_proposed_2016,gomide_dengue_2011} \\ 
\rowcolor{white}\multirow{3}{*}{\cellcolor[rgb]{0.961,0.961,0.961}} & ~ & ~ \\[-5ex]
\multirow{-2}{*}{\cellcolor[rgb]{0.961,0.961,0.961}} & {\cellcolor[rgb]{0.961,0.961,0.961}} \textbf{Ebola} & {\cellcolor[rgb]{0.961,0.961,0.961}} \textbf{Twitter} & {\cellcolor[rgb]{0.961,0.961,0.961}} \cite{marc_how_2020,yom-tov_ebola_2015} \\ 
\multirow{-2}{*}{\cellcolor[rgb]{0.961,0.961,0.961}} & {\cellcolor[rgb]{0.961,0.961,0.961}} & {\cellcolor[rgb]{0.961,0.961,0.961}}\textbf{Weibo}  & {\cellcolor[rgb]{0.961,0.961,0.961}}\cite{feng_weibo_2016}  \\ 
\rowcolor{white}\multirow{3}{*}{\cellcolor[rgb]{0.961,0.961,0.961}} & ~ & ~ \\[-5ex]
\multirow{-2}{*}{\cellcolor[rgb]{0.961,0.961,0.961}} & {\cellcolor[rgb]{0.961,0.961,0.961}} \textbf{H1N1/Swine flu} & {\cellcolor[rgb]{0.961,0.961,0.961}} \textbf{Twitter} & {\cellcolor[rgb]{0.961,0.961,0.961}} \cite{szomszor__2010,de_quincey_early_2009}  \\ 
\rowcolor{white}\multirow{3}{*}{\cellcolor[rgb]{0.961,0.961,0.961}} & ~ & ~ \\[-5ex]
\multirow{-2}{*}{\cellcolor[rgb]{0.961,0.961,0.961}} & {\cellcolor[rgb]{0.961,0.961,0.961}} \textbf{Influenza/Flu} & {\cellcolor[rgb]{0.961,0.961,0.961}} \textbf{Twitter} & {\cellcolor[rgb]{0.961,0.961,0.961}}\cite{stilo_predicting_2014,hwang_spatiotemporal_2013,chon_modeling_2015, lampos_enhancing_2017,lampos_flu_2010,talvis_real-time_2014,byrd_mining_2016,comito_improving_2018,kostkova__2014}  \\ 
\multirow{-2}{*}{\cellcolor[rgb]{0.961,0.961,0.961}} & {\cellcolor[rgb]{0.961,0.961,0.961}} & {\cellcolor[rgb]{0.961,0.961,0.961}}\textbf{Sina Weibo, Tancent Weibo} & {\cellcolor[rgb]{0.961,0.961,0.961}}\cite{sun_real_2014,zhang_social_2015} \\ 
\rowcolor{white}\multirow{3}{*}{\cellcolor[rgb]{0.961,0.961,0.961}} & ~ & ~ \\[-5ex]
\multirow{-2}{*}{\cellcolor[rgb]{0.961,0.961,0.961}} & {\cellcolor[rgb]{0.961,0.961,0.961}} \textbf{Zika} & {\cellcolor[rgb]{0.961,0.961,0.961}} \textbf{Twitter} & {\cellcolor[rgb]{0.961,0.961,0.961}}\cite{masri_use_2019,abouzahra_twitter_2021}  \\ 
\multirow{-2}{*}{\cellcolor[rgb]{0.961,0.961,0.961}} & {\cellcolor[rgb]{0.961,0.961,0.961}} & {\cellcolor[rgb]{0.961,0.961,0.961}} \textbf{Reddit} & {\cellcolor[rgb]{0.961,0.961,0.961}} \cite{gui_managing_2017}  \\ 
\multirow{-2}{*}{\cellcolor[rgb]{0.961,0.961,0.961}} & {\cellcolor[rgb]{0.961,0.961,0.961}} \textbf{MERS}  & {\cellcolor[rgb]{0.961,0.961,0.961}} \textbf{Twitter} & {\cellcolor[rgb]{0.961,0.961,0.961}}\cite{chaudhary_use_2017}  \\ 
\multirow{-2}{*}{\cellcolor[rgb]{0.961,0.961,0.961}} & {\cellcolor[rgb]{0.961,0.961,0.961}} & {\cellcolor[rgb]{0.961,0.961,0.961}} \textbf{Facebook} & {\cellcolor[rgb]{0.961,0.961,0.961}}\cite{chaudhary_use_2017}  \\ 
\rowcolor{white}\multirow{3}{*}{\cellcolor[rgb]{0.961,0.961,0.961}} & ~ & ~ \\[-5ex]
\multirow{-2}{*}{\cellcolor[rgb]{0.961,0.961,0.961}} & {\cellcolor[rgb]{0.961,0.961,0.961}} \textbf{Multiple Epidemics}  & {\cellcolor[rgb]{0.961,0.961,0.961}} \textbf{Twitter} & {\cellcolor[rgb]{0.961,0.961,0.961}}\cite{romano_challenges_2016,ji_epidemic_2012,chaudhary_use_2017}  \\ 
\multirow{-2}{*}{\cellcolor[rgb]{0.961,0.961,0.961}} & {\cellcolor[rgb]{0.961,0.961,0.961}} & {\cellcolor[rgb]{0.961,0.961,0.961}} \textbf{Facebook} & {\cellcolor[rgb]{0.961,0.961,0.961}} \cite{chaudhary_use_2017}  \\ 
\rowcolor{white}\multirow{3}{*}~ & ~ & ~ \\[-5ex]
\multirow{3}{*}{\cellcolor[rgb]{0.851,0.922,0.922}}{\textbf{ML Classification}}  & {\cellcolor[rgb]{0.851,0.922,0.922}} \textbf{Dengue Fever} & {\cellcolor[rgb]{0.851,0.922,0.922}} \textbf{Twitter} & {\cellcolor[rgb]{0.851,0.922,0.922}} \cite{souza_evolutionary_2014,souza_infection_2016} \\ 
\multirow{-2}{*}{\cellcolor[rgb]{0.851,0.922,0.922}} & {\cellcolor[rgb]{0.851,0.922,0.922}} \textbf{Influenza/Flu} & {\cellcolor[rgb]{0.851,0.922,0.922}}\textbf{Twitter} & {\cellcolor[rgb]{0.851,0.922,0.922}}\cite{lamb_separating_2013,zadeh_social_2019,zhang_predicting_2017,wakamiya_after_2016,aramaki_twitter_2011,zhang_detecting_2014,achrekar_online_2012,culotta_towards_2010,sun_hybrid_2015,xiao_detecting_2016} \\
\multirow{-2}{*}{\cellcolor[rgb]{0.851,0.922,0.922}} & {\cellcolor[rgb]{0.851,0.922,0.922}} & {\cellcolor[rgb]{0.851,0.922,0.922}}\textbf{Facebook} & {\cellcolor[rgb]{0.851,0.922,0.922}} \cite{achrekar_online_2012}  \\
\multirow{-2}{*}{\cellcolor[rgb]{0.851,0.922,0.922}} & {\cellcolor[rgb]{0.851,0.922,0.922}} & {\cellcolor[rgb]{0.851,0.922,0.922}}\textbf{Sina Weibo, Tancent Weibo} & {\cellcolor[rgb]{0.851,0.922,0.922}} \cite{zhang_predicting_2017,sun_hybrid_2015,sun_real_2014}  \\
\rowcolor{white}\multirow{3}{*}{\cellcolor[rgb]{0.851,0.922,0.922}} & ~ & ~ \\[-5ex]
\multirow{-2}{*}{\cellcolor[rgb]{0.851,0.922,0.922}} & {\cellcolor[rgb]{0.851,0.922,0.922}} \textbf{H1N1/Swine Flu} & {\cellcolor[rgb]{0.851,0.922,0.922}}\textbf{Twitter} & {\cellcolor[rgb]{0.851,0.922,0.922}}\cite{collier_omg_2011} \\
\rowcolor{white}\multirow{3}{*}{\cellcolor[rgb]{0.851,0.922,0.922}} & ~ & ~ \\[-5ex]
\multirow{-2}{*}{\cellcolor[rgb]{0.851,0.922,0.922}} & {\cellcolor[rgb]{0.851,0.922,0.922}} \textbf{MERS} & {\cellcolor[rgb]{0.851,0.922,0.922}} \textbf{Twitter} & {\cellcolor[rgb]{0.851,0.922,0.922}} \cite{rudra_classifying_2017,rudra_classifying_2018,ghosh_identifying_2019}  \\ 
\rowcolor{white}\multirow{3}{*}{\cellcolor[rgb]{0.851,0.922,0.922}} & ~ & ~ \\[-5ex]
\multirow{-2}{*}{\cellcolor[rgb]{0.851,0.922,0.922}} & {\cellcolor[rgb]{0.851,0.922,0.922}} \textbf{Ebola} & {\cellcolor[rgb]{0.851,0.922,0.922}} \textbf{Twitter} & {\cellcolor[rgb]{0.851,0.922,0.922}} \cite{khatua_tale_2019} \\
\rowcolor{white}\multirow{3}{*}{\cellcolor[rgb]{0.851,0.922,0.922}} & ~ & ~ \\[-5ex]
\multirow{-2}{*}{\cellcolor[rgb]{0.851,0.922,0.922}} & {\cellcolor[rgb]{0.851,0.922,0.922}} \textbf{Zika} & {\cellcolor[rgb]{0.851,0.922,0.922}} \textbf{Twitter} & {\cellcolor[rgb]{0.851,0.922,0.922}} \cite{khatua_tale_2019} \\ 
\multirow{-2}{*}{\cellcolor[rgb]{0.851,0.922,0.922}} & {\cellcolor[rgb]{0.851,0.922,0.922}} \textbf{Multiple Epidemics} & {\cellcolor[rgb]{0.851,0.922,0.922}} \textbf{Twitter} & {\cellcolor[rgb]{0.851,0.922,0.922}} \cite{sousa_vazadengue_2018,hong_social_2018,swain_analysis_2017,rudra_classifying_2017,rudra_classifying_2018,ghosh_identifying_2019,jain_effective_2018}  \\ 
\rowcolor{white}\multirow{3}{*}~ & ~ & ~ \\[-5ex]
\multirow{4}{*}{\cellcolor[rgb]{0.961,0.961,0.961}}{\textbf{DL Classification}} & {\cellcolor[rgb]{0.961,0.961,0.961}} \textbf{Covid-19} & {\cellcolor[rgb]{0.961,0.961,0.961}} \textbf{Twitter} & {\cellcolor[rgb]{0.961,0.961,0.961}} \cite{lwowski_covid-19_2020,mundotiya_nlprl_2020,cornelius_covid-19_2020,rao_retweets_2020} \\ 
\multirow{-2}{*}{\cellcolor[rgb]{0.961,0.961,0.961}} & {\cellcolor[rgb]{0.961,0.961,0.961}} \textbf{Ebola} & {\cellcolor[rgb]{0.961,0.961,0.961}} \textbf{Twitter} & {\cellcolor[rgb]{0.961,0.961,0.961}} \cite{khatua_tale_2019} \\ 
\multirow{-2}{*}{\cellcolor[rgb]{0.961,0.961,0.961}} & {\cellcolor[rgb]{0.961,0.961,0.961}} \textbf{Zika} & {\cellcolor[rgb]{0.961,0.961,0.961}} \textbf{Twitter} & {\cellcolor[rgb]{0.961,0.961,0.961}} \cite{khatua_tale_2019} \\ 
\rowcolor{white}\multirow{3}{*}{\cellcolor[rgb]{0.961,0.961,0.961}} & ~ & ~ \\[-5ex]
\multirow{-2}{*}{\cellcolor[rgb]{0.961,0.961,0.961}} & {\cellcolor[rgb]{0.961,0.961,0.961}} \textbf{Influenza/Flu} & {\cellcolor[rgb]{0.961,0.961,0.961}} \textbf{Twitter} & {\cellcolor[rgb]{0.961,0.961,0.961}} \cite{wang_using_2017} \\ 
\rowcolor{white}\multirow{3}{*}{\cellcolor[rgb]{0.961,0.961,0.961}} & ~ & ~ \\[-5ex]
\multirow{-2}{*}{\cellcolor[rgb]{0.961,0.961,0.961}} & {\cellcolor[rgb]{0.961,0.961,0.961}} \textbf{Multiple Epidemics} & {\cellcolor[rgb]{0.961,0.961,0.961}} \textbf{Twitter} & {\cellcolor[rgb]{0.961,0.961,0.961}} \cite{amin_detecting_2020} \\ 
\rowcolor{white}\multirow{3}{*}~ & ~ & ~ \\[-5ex]
{\cellcolor[rgb]{0.851,0.922,0.922}}\textbf{Mathematical Modeling} & {\cellcolor[rgb]{0.851,0.922,0.922}} \textbf{Covid-19} & {\cellcolor[rgb]{0.851,0.922,0.922}} \textbf{WeChat}  & {\cellcolor[rgb]{0.851,0.922,0.922}} \cite{nie2021dynamical} \\ 
\rowcolor{white}\multirow{3}{*}{\cellcolor[rgb]{0.851,0.922,0.922}} & ~ & ~ \\[-5ex]
\multirow{-2}{*}{\cellcolor[rgb]{0.851,0.922,0.922}} & {\cellcolor[rgb]{0.851,0.922,0.922}} \textbf{H1N1/Swine Flu} & {\cellcolor[rgb]{0.851,0.922,0.922}} \textbf{Twitter}  & {\cellcolor[rgb]{0.851,0.922,0.922}} \cite{huo_modeling_2016,zhao_online_2020,huo_modeling_2020} \\ 
\rowcolor{white}\multirow{3}{*}~ & ~ & ~ \\[-5ex]
{\cellcolor[rgb]{0.961,0.961,0.961}}\textbf{SNA}   & {\cellcolor[rgb]{0.961,0.961,0.961}} \textbf{Covid-19} & {\cellcolor[rgb]{0.961,0.961,0.961}} \textbf{Twitter} & {\cellcolor[rgb]{0.961,0.961,0.961}} \cite{spurlock_predicting_2020,haupt2021characterizing}\\ 
\rowcolor{white}\multirow{3}{*}{\cellcolor[rgb]{0.961,0.961,0.961}} & ~ & ~ \\[-5ex]
\multirow{-2}{*}{\cellcolor[rgb]{0.961,0.961,0.961}} & {\cellcolor[rgb]{0.961,0.961,0.961}} \textbf{Influenza/Flu}  & {\cellcolor[rgb]{0.961,0.961,0.961}} \textbf{Facebook} & {\cellcolor[rgb]{0.961,0.961,0.961}} \cite{martin_leveraging_2011} \\  
\rowcolor{white}\multirow{3}{*}{\cellcolor[rgb]{0.961,0.961,0.961}} & ~ & ~ \\[-5ex]
\multirow{-2}{*}{\cellcolor[rgb]{0.961,0.961,0.961}} & {\cellcolor[rgb]{0.961,0.961,0.961}} \textbf{Multiple Epidemics} & {\cellcolor[rgb]{0.961,0.961,0.961}} \textbf{Twitter} & {\cellcolor[rgb]{0.961,0.961,0.961}}  \cite{bagavathi_social_2018}  \\
\rowcolor{white}\multirow{3}{*}~ & ~ & ~ \\[-5ex]
\multirow{4}{*}{\cellcolor[rgb]{0.851,0.922,0.922}}{\cellcolor[rgb]{0.851,0.922,0.922}}\textbf{Topic Modeling} & {\cellcolor[rgb]{0.851,0.922,0.922}} \textbf{Covid-19} & {\cellcolor[rgb]{0.851,0.922,0.922}} \textbf{Twitter} & {\cellcolor[rgb]{0.851,0.922,0.922}}\cite{cornelius_covid-19_2020,zheng_uncovering_2020,haupt2021characterizing,shin2020risk} \\ 
\multirow{-2}{*}{\cellcolor[rgb]{0.851,0.922,0.922}} & {\cellcolor[rgb]{0.851,0.922,0.922}} \textbf{Dengue Fever} & {\cellcolor[rgb]{0.851,0.922,0.922}} \textbf{Twitter} & {\cellcolor[rgb]{0.851,0.922,0.922}} \cite{missier_tracking_2016} \\ 
\rowcolor{white}\multirow{3}{*}{\cellcolor[rgb]{0.851,0.922,0.922}} & ~ & ~ \\[-5ex]
\multirow{-2}{*}{\cellcolor[rgb]{0.851,0.922,0.922}} & {\cellcolor[rgb]{0.851,0.922,0.922}} \textbf{Ebola} & {\cellcolor[rgb]{0.851,0.922,0.922}} \textbf{Twitter} & {\cellcolor[rgb]{0.851,0.922,0.922}} \cite{odlum_what_2015} \\ 
\multirow{-2}{*}{\cellcolor[rgb]{0.851,0.922,0.922}} & {\cellcolor[rgb]{0.851,0.922,0.922}} \textbf{Influenza/Flu} & {\cellcolor[rgb]{0.851,0.922,0.922}} \textbf{Twitter} & {\cellcolor[rgb]{0.851,0.922,0.922}} \cite{chen_flu_2014} \\
\multirow{-2}{*}{\cellcolor[rgb]{0.851,0.922,0.922}} & {\cellcolor[rgb]{0.851,0.922,0.922}} & {\cellcolor[rgb]{0.851,0.922,0.922}} \textbf{Weibo} & {\cellcolor[rgb]{0.851,0.922,0.922}}\cite{fu_detect_2014} \\
\rowcolor{white}\multirow{3}{*}{\cellcolor[rgb]{0.851,0.922,0.922}} & ~ & ~ \\[-5ex]
\multirow{-2}{*}{\cellcolor[rgb]{0.851,0.922,0.922}} & {\cellcolor[rgb]{0.851,0.922,0.922}} \textbf{Zika} & {\cellcolor[rgb]{0.851,0.922,0.922}} \textbf{Twitter} & {\cellcolor[rgb]{0.851,0.922,0.922}} \cite{nolasco_subevents_2019}  \\
\rowcolor{white}\multirow{3}{*}{\cellcolor[rgb]{0.851,0.922,0.922}} & ~ & ~ \\[-5ex]
\multirow{-2}{*}{\cellcolor[rgb]{0.851,0.922,0.922}} & {\cellcolor[rgb]{0.851,0.922,0.922}} \textbf{Multiple Epidemics} & {\cellcolor[rgb]{0.851,0.922,0.922}} \textbf{Twitter} & {\cellcolor[rgb]{0.851,0.922,0.922}} \cite{diaz-aviles_tracking_2012} \\
\end{longtable}
\end{small}

\paragraph{\textbf{b. Public opinion understanding}}

Several methods were used in the literature to extract and analyze public opinions expressed on social media. 
The work presented in \cite{chetty_2019-ncov_2020} examined Twitter data and analyzed it to understand the various discussions taking place online in relation to Covid-19. Results indicate the association of the words "coronavirus" and "china". The presence of negativity in Covid-19 related discussions on social media is noted by the authors.
The authors of \cite{shanthakumar_understanding_2020} identified and racked Covid-19 related hashtags, and as a result collected a total of 530,206 tweets, which they grouped into six categories: General Covid, Quarantine, Panic buying, School closures, Lockdowns, Frustration and hope. 
Both general and hashtag-group-specific linguistic content analysis was performed with the aim of analyzing the socio-economic disruption caused by COVID-19 in the United States, understanding the chain events taking place during the pandemic, and to draw lessons on what could be avoided in the future. Temporal evolution results revealed increased calls for social distancing, quarantining, working from home to limit the spread of the
disease, followed by hashtag expressions of anger at individuals refusing to respect Covid-19 protocols. Words such as family, life, health and death were found to be common across hashtag groups, while mentions to mental health was observed as a mention to possible psychological consequences of social isolation. Solidarity and gratitude for essential workers and gratitude were also noted in several mentions.
The study presented in \cite{wang_examining_2021} analyzed 13,598 pandemic relevant tweets from various agencies and stakeholders in the United States, and manually annotated them following 16 categories of topics (strategies, closures, openings, resource provision, employment, etc.). A network analysis examining risk and crisis communications of government agencies and stakeholders was performed to illustrate weekly changes and network communities. The results reported in this study include an increased level of connectivity and agency coordination during the early-stage response.  
Similarly, \cite{niknam_covid-19_2021} performed a thematic content analysis To characterize the representation of public health information related to COVID-19 posted on Instagram in 2020. As a result, 23 themes were identified, including statistics, prevention, hygiene, diagnosis, politics, world news, etc. Qualitative document and content analysis of social media platforms was conducted in \cite{olcer_lay_2020} and revealed that the uncertainty related to Covid-19, and the range of "noise" related to it on social media, along with its socio-economic consequences, could lead to individuals' disregard for sanitary and government recommendations.  
Similar content analysis was performed in the context of Ebola, Zika and Influenza. Health organizations' social media posts were analyzed in \cite{guidry_ebola_2017}, and were found to be highly effective when incorporation visuals. Public response was found to be more affected by these communications when they acknowledged the concerns and fear of the community. The evolution of Ebola related online blame was analyzed in \cite{roy_ebola_2020}. Findings highlighted the direction of blame towards the affected populations as well as figures with whom social media users had pre-existing political frustrations.
Content analysis was performed in \cite{gui_multidimensional_2018} to manually analyze public Discourse around Zika related risks, and in \cite{lehmann_qualitative_2013} to inspect social media coverage related to Influenza (Flu and Swine Flu) vaccinations.
Independent manual annotation aiming to inspect how Arab Twitter users perceived Covid-19 was performed in \cite{essam_how_2021}, and generated multiple categories including: conspiracy, economy, mortality, origin, outbreak, politics,etc. Social media analytics were found to be an efficient approach to capture the attitudes and perceptions of the public during Covid-19 in \cite{yigitcanlar_how_2020,xia_outlier_2020}.
Sentiment analysis was performed to measure public health concerns \cite{ji_twitter_2015}, and to analyse polarity over time, as well as to identify stance towards two specific pandemic policies regarding social distancing and wearing face masks \cite{wang_public_2020}. Using a lexicon-based analysis and a multiple regression model, \cite{huang_how_2020} found that fear and collectivism can predict people's preventive intention in the context of Covid-19. In \cite{munoz_data-driven_2020}, VADER\footnote{\url{https://github.com/cjhutto/vaderSentiment}} (Valence Aware Dictionary and Sentiment Reasoner), a lexicon and rule-based social media sentiment analysis tool was tool on a Twitter dataset collected before the Covid-19 pandemic (December 2019), when Covid-19 was declared a pandemic by the WHO (March 2020), and at the beginning of protocol relaxation in a few countries (May 2020), in order to inspect the presence and escalation of negative sentiments towards China. 

Conventional machine learning models and deep learning models were also used in several works aiming to understand public opinion in epidemics.

Sentiment analysis was performed using Naive Bayes and Support Vector Machine to track and understand public reactions during Covid-19
investigate topics and sentiment \cite{addawood_tracking_2020}. Clustering and classification were incorporated in a proposed methodology in \cite{satu2021tclustvid} for the extraction of significant topics related to the pandemic.  
Supervised learning methods such as support vector machines (SVM), naive Bayes (NB), and random forest (RF) were used in \cite{li_characterizing_2020} to classify social media content into several types (Caution and advice, notifications or measures, donation, etc.) of situational information related to the pandemic. \cite{mandal_predicting_2018} classified Zika related tweets into various prevention categories as proposed by CDC. A logistic regression model achieved a 71\% accuracy in this classification.

BERTmoticon was proposed as a fine-tuned BERT (Bidirectional Encoder Representations from Transformers) model in \cite{stoikos_multilingual_2020} for multilingual emoticon prediction in relation to Covid-19. This work reported in its findings that sadness spiked when the World Health Organization declared Covid-19 a pandemic, and that anger and disgust spiked after the death toll surpassed the hundred thousands in the United States. 
A pre-trained GPT (Generative pre-trained transformer) model was customized with Covid-19 tweets was applied to reveal insights into the biases and opinions of the users \cite{feldman_analyzing_2021}. A universal language model for the Moroccan dialect was built in \cite{ghanem2021real} and fine-tuned using a collected Covid-19 dataset to perform topic modeling, emotion recognition and polar sentiment analysis, and understand the Moroccan population's feelings towards the pandemic and the government's response to it. LSTM (Long Short Term Memory), BERT (Bidirectional Encoder Representations from Transformers) and ERNIE (Enhanced Language Representation with Informative Entities) were used in \cite{lyu_sentiment_2020} to analyze the evolution of sentiments in the face of Covid-19's public health crisis. 
\cite{shah_sentiment_2020} performed Latent Semantic Analysis (LSA) and Latent Dirichlet Allocation (LDA) to mine opinions on Twitter related to the hashtag \#IndiaFightsCorona. Sentiment analysis revealed that positive sentiments outweighed negative ones. 
Term-frequency analysis is adopted in \cite{de_santis_infoveillance_2020} for building a topic graph where emerging topics are suitably selected, while k-means algorithm and other conventional machine learning models (Logistic Regression, Support Vector Classification, and Naïve Bayes) are used in \cite{alsudias_covid-19_2020} to identify Covid-19 related topics.
Topic detection and sentiment analysis were conducted in \cite{garcia_topic_2021,massaro_non-pharmaceutical_2021,shi_automated_2021,chen_covid-19_2020,zhu_analysis_2020} for the purposes of opinion mining, concern exploration, and public opinion analysis in the context of epidemics. \\

\textbf{\textit{Tab. \ref{table:rq1surveillance}}} and \textbf{\textit{Tab. \ref{table:rq1opinion}}} summarize the methods, epidemics and social media used in studies pertaining epidemic forecasting and prediction and public opinion understanding.
\renewcommand{\arraystretch}{2.25}
\begin{small}
\begin{longtable}{p{4.5cm}p{3cm}p{3cm}p{2cm}}
\caption{\textbf{Summary of methodologies used in studies addressing research question 1: Public Opinion Understanding}}\\
\label{table:rq1opinion}\\
\rowcolor[rgb]{0.604,0.8,0.8}\textbf{Method} & \textbf{Epidemic Studied} & \textbf{Social Media Used} & \textbf{References} \\
\multirow{3}{*}~ & ~ & ~ \\[-5ex]
\multirow{5}{*}{\cellcolor[rgb]{0.961,0.961,0.961}}{\textbf{Content Analysis}} & {\cellcolor[rgb]{0.961,0.961,0.961}} \multirow{3}{*}{\cellcolor[rgb]{0.961,0.961,0.961}}\textbf{Covid-19} & {\cellcolor[rgb]{0.961,0.961,0.961}} \textbf{Twitter} & {\cellcolor[rgb]{0.961,0.961,0.961}} \cite{chetty_2019-ncov_2020,shanthakumar_understanding_2020,wang_examining_2021} \\ 
\multirow{-2}{*}{\cellcolor[rgb]{0.961,0.961,0.961}} & {\cellcolor[rgb]{0.961,0.961,0.961}} & {\cellcolor[rgb]{0.961,0.961,0.961}} \textbf{Instagram} & {\cellcolor[rgb]{0.961,0.961,0.961}} \cite{niknam_covid-19_2021} \\ 
\multirow{-2}{*}{\cellcolor[rgb]{0.961,0.961,0.961}} & {\cellcolor[rgb]{0.961,0.961,0.961}}& {\cellcolor[rgb]{0.961,0.961,0.961}}\textbf{Reddit} & {\cellcolor[rgb]{0.961,0.961,0.961}} \cite{olcer_lay_2020} \\ 
\rowcolor{white}\multirow{3}{*}{\cellcolor[rgb]{0.961,0.961,0.961}} & ~ & ~ \\[-5ex]
\multirow{-2}{*}{\cellcolor[rgb]{0.961,0.961,0.961}} & {\cellcolor[rgb]{0.961,0.961,0.961}} \textbf{Ebola} & {\cellcolor[rgb]{0.961,0.961,0.961}}\textbf{Twitter} & {\cellcolor[rgb]{0.961,0.961,0.961}} \cite{guidry_ebola_2017,roy_ebola_2020} \\ 
\multirow{-2}{*}{\cellcolor[rgb]{0.961,0.961,0.961}} & {\cellcolor[rgb]{0.961,0.961,0.961}}& {\cellcolor[rgb]{0.961,0.961,0.961}}\textbf{Facebook} & {\cellcolor[rgb]{0.961,0.961,0.961}} \cite{roy_ebola_2020} \\ 
\multirow{-2}{*}{\cellcolor[rgb]{0.961,0.961,0.961}} & {\cellcolor[rgb]{0.961,0.961,0.961}}& {\cellcolor[rgb]{0.961,0.961,0.961}}\textbf{Instagram} & {\cellcolor[rgb]{0.961,0.961,0.961}} \cite{guidry_ebola_2017} \\ 
\rowcolor{white}\multirow{3}{*}{\cellcolor[rgb]{0.961,0.961,0.961}} & ~ & ~ \\[-5ex]
\multirow{-2}{*}{\cellcolor[rgb]{0.961,0.961,0.961}} & {\cellcolor[rgb]{0.961,0.961,0.961}} \textbf{Zika}  & {\cellcolor[rgb]{0.961,0.961,0.961}} \textbf{Reddit} & {\cellcolor[rgb]{0.961,0.961,0.961}} \cite{gui_multidimensional_2018}  \\ 
\rowcolor{white}\multirow{3}{*}{\cellcolor[rgb]{0.961,0.961,0.961}} & ~ & ~ \\[-5ex]
\multirow{-2}{*}{\cellcolor[rgb]{0.961,0.961,0.961}} & {\cellcolor[rgb]{0.961,0.961,0.961}} \textbf{Influenza/Flu}  & {\cellcolor[rgb]{0.961,0.961,0.961}} \textbf{Twitter} & {\cellcolor[rgb]{0.961,0.961,0.961}} \cite{lehmann_qualitative_2013} \\
\multirow{-2}{*}{\cellcolor[rgb]{0.961,0.961,0.961}} & {\cellcolor[rgb]{0.961,0.961,0.961}} & {\cellcolor[rgb]{0.961,0.961,0.961}} \textbf{Facebook} & {\cellcolor[rgb]{0.961,0.961,0.961}} \cite{lehmann_qualitative_2013} \\
\rowcolor{white}\multirow{3}{*}{\cellcolor[rgb]{0.961,0.961,0.961}} & ~ & ~ \\[-5ex]
\multirow{-2}{*}{\cellcolor[rgb]{0.961,0.961,0.961}} & {\cellcolor[rgb]{0.961,0.961,0.961}} \textbf{H1N1/Swine Flu}  & {\cellcolor[rgb]{0.961,0.961,0.961}} \textbf{Twitter} & {\cellcolor[rgb]{0.961,0.961,0.961}} \cite{lehmann_qualitative_2013} \\
\multirow{-2}{*}{\cellcolor[rgb]{0.961,0.961,0.961}} & {\cellcolor[rgb]{0.961,0.961,0.961}} & {\cellcolor[rgb]{0.961,0.961,0.961}} \textbf{Facebook} & {\cellcolor[rgb]{0.961,0.961,0.961}} \cite{lehmann_qualitative_2013} \\
\multirow{3}{*}~ & ~ & ~ \\[-5ex]
\multirow{3}{*}{\cellcolor[rgb]{0.851,0.922,0.922}}{\textbf{Dictionary-based Classification}} & {\cellcolor[rgb]{0.851,0.922,0.922}}\textbf{Covid19} & {\cellcolor[rgb]{0.851,0.922,0.922}}\textbf{Twitter} & {\cellcolor[rgb]{0.851,0.922,0.922}}\cite{essam_how_2021,yigitcanlar_how_2020,xia_outlier_2020,ji_twitter_2015,wang_public_2020,munoz_data-driven_2020} \\ 
\multirow{-2}{*}{\cellcolor[rgb]{0.851,0.922,0.922}} & {\cellcolor[rgb]{0.851,0.922,0.922}} & {\cellcolor[rgb]{0.851,0.922,0.922}}\textbf{Reddit} & {\cellcolor[rgb]{0.851,0.922,0.922}} \cite{wang_public_2020} \\ 
\rowcolor{white}\multirow{3}{*}{\cellcolor[rgb]{0.851,0.922,0.922}} & ~ & ~ \\[-5ex]
\multirow{-2}{*}{\cellcolor[rgb]{0.851,0.922,0.922}} & {\cellcolor[rgb]{0.851,0.922,0.922}} \textbf{Multiple Epidemics} & {\cellcolor[rgb]{0.851,0.922,0.922}} \textbf{Weibo} & {\cellcolor[rgb]{0.851,0.922,0.922}} \cite{huang_how_2020} \\ 
\multirow{3}{*}~ & ~ & ~ \\[-5ex]
\multirow{2}{*}{\cellcolor[rgb]{0.961,0.961,0.961}}{\textbf{ML Classification}}  & {\cellcolor[rgb]{0.961,0.961,0.961}}\textbf{Covid19} & {\cellcolor[rgb]{0.961,0.961,0.961}}\textbf{Twitter} & {\cellcolor[rgb]{0.961,0.961,0.961}}\cite{addawood_tracking_2020,satu2021tclustvid,shah_sentiment_2020} \\ 
\multirow{-2}{*}{\cellcolor[rgb]{0.961,0.961,0.961}} & {\cellcolor[rgb]{0.961,0.961,0.961}} & {\cellcolor[rgb]{0.961,0.961,0.961}} \textbf{Weibo} & {\cellcolor[rgb]{0.961,0.961,0.961}} \cite{li_characterizing_2020}  \\
\rowcolor{white}\multirow{3}{*}{\cellcolor[rgb]{0.961,0.961,0.961}} & ~ & ~ \\[-5ex]
\multirow{-2}{*}{\cellcolor[rgb]{0.961,0.961,0.961}} & {\cellcolor[rgb]{0.961,0.961,0.961}} \textbf{Zika} & {\cellcolor[rgb]{0.961,0.961,0.961}} \textbf{Twitter} & {\cellcolor[rgb]{0.961,0.961,0.961}} \cite{mandal_predicting_2018}  \\
\multirow{3}{*}~ & ~ & ~ \\[-5ex]
{\cellcolor[rgb]{0.851,0.922,0.922}}\textbf{DL Classification} & {\cellcolor[rgb]{0.851,0.922,0.922}} \textbf{Covid19} & {\cellcolor[rgb]{0.851,0.922,0.922}} \textbf{Twitter} & {\cellcolor[rgb]{0.851,0.922,0.922}} \cite{stoikos_multilingual_2020,feldman_analyzing_2021,ghanem2021real}  \\ 
\multirow{-2}{*}{\cellcolor[rgb]{0.851,0.922,0.922}} & {\cellcolor[rgb]{0.851,0.922,0.922}} & {\cellcolor[rgb]{0.851,0.922,0.922}} \textbf{Weibo} & {\cellcolor[rgb]{0.851,0.922,0.922}} \cite{lyu_sentiment_2020} \\
\multirow{3}{*}~ & ~ & ~ \\[-5ex]
{\cellcolor[rgb]{0.961,0.961,0.961}}\textbf{Topic Modeling}   & {\cellcolor[rgb]{0.961,0.961,0.961}} \textbf{Covid19} & {\cellcolor[rgb]{0.961,0.961,0.961}} \textbf{Twitter} & {\cellcolor[rgb]{0.961,0.961,0.961}} \cite{de_santis_infoveillance_2020,alsudias_covid-19_2020,garcia_topic_2021,massaro_non-pharmaceutical_2021,shi_automated_2021,ghanem2021real}
\\ 
\multirow{-2}{*}{\cellcolor[rgb]{0.961,0.961,0.961}} & {\cellcolor[rgb]{0.961,0.961,0.961}} & {\cellcolor[rgb]{0.961,0.961,0.961}} \textbf{Weibo} & {\cellcolor[rgb]{0.961,0.961,0.961}} \cite{chen_covid-19_2020,zhu_analysis_2020} \\
\end{longtable}
\end{small}

\subsubsection{RQ2: Can social media be used for misinformation management during epidemics?}

Misinformation, or 'fake news', has become a social phenomenon and has received increased attention in the past few years. Although the 'fake news' term has been around since since the 1890s (\cite{machete2020use}), the emergence and exponential rise in popularity of social media platforms has brought the term to the \textit{front page}. Fake news can fall into multiple categories depending on the intent and form it takes \cite{machete2020use}. For instance, fake news can be false information and rumor fabrication (e.g., celebrity gossip), hoaxes (e.g., doomsday 2012), conspiracy theories (Q-Anon) and satire (e.g., The Onion). The intent can range from deception for the purposes of monetary or personal gain, to satirizing real news.

One main theme was identified in the selected papers addressing how social media can be used in misinformation management during epidemics, namely: (a) Misinformation detection and characterization. Three subsequent sub-themes were identified based on the scope of selected literature, namely: (a.1) Fake news identification, (a.2) Fake news characterization, and (a.3) Information distortion and conspiracy theories.
Methods used in each of these themes were classified into groups depending on the aim and results of the study.
\paragraph{\textbf{a. Misinformation detection and characterization}}

It was observed that the selected literature focused on the inspection of news or claims shared on social media, with the aim of classifying them based on their trustworthiness. Several methods were used to analyze social media content and detect detect misleading information, from expert annotation to deep learning models and social network analysis. 
While some papers focused on technical approaches to the detection of fake news, other studies tried to identify various characteristics related to the source or propagation of fake news.

\textit{\textbf{a.1. Fake news Identification}}

In the work presented in \cite{abdelminaam2021coaid}, a deep neural network is proposed for the identification of fake news using modified LSTM (Long Short Term Memory) and modified GRU (Gated Recurrent Unit), each with one to three layers. Additionally, six conventional machine learning classification models were used for the same task:  decision trees, logistic regression, k nearest neighbors, random forests, support vector machines, and naïve Bayes (NB). 4 Twitter fake news datasets were used in this study: CoAID (COVID-19 heAlthcare mIsinformation
Dataset) \cite{cui2020coaid}, Kaggle's disaster topic \cite{sarin2020convgrutext}, FakeNewsNet's PolitiFact and gossipcop \cite{shu2018fakenewsnet}. The feature analysis of the ML approach is based on TF-IDF and Ngrams, while the deep learning approach depends on word embedding. Both approaches are optimized using grid search and Keras tuning, respectively. For the Covid-19 dataset, the best testing results are obtained by LSTM (two layers), with an accuracy of 98.6\%, a precision of 98.55\%, a recall of 98.6\% and an F1-score of 98.5\%. 

The authors of \cite{madani2021using} used conventional machine learning models (Logistic Regression, Decision Tree, Random Forest, Naive Bayes, SVM, Gradient Boosting) and deep learning models (Multilayer perceptron) to detect Covid-19 related fake news in tweets.
This work used two Twitter datasets: FakeNewsNet's PolitiFact and Gossipcop \cite{shu2018fakenewsnet}, and a disaster dataset \cite{cherylfake} including "Las Vegas shooting, 2017" and "Hurricane Harvey, 2017". A dataset of Covid-19 Moroccan tweets was collected in their publication language (Arabic, Spanish, French, English), and translated to English for the classification task. In addition to the text of the tweet, the authors used features such as: source, retweet count, user name, followers count and sentiment of the tweet (positive, negative, neutral). The results reported in this work highlight that the Random Forest model outperformed all other models, with an accuracy of 78\%, a recall of 100\%, a precision of 85\%, and an F1-score of 83\%. Additionally, the authors conclude that from the 2000 Covid-19 Moroccan tweets collected, 37\% represent fake news. 

Similarly, \cite{maakoul2020towards} used conventional machine learning in the form of logistic regression to detect fake news from a dataset of Facebook comments. This work, however, is limited by the dataset used consisting only of 80 comments, and the use of only one model. The authors note the existence of false positives in their results and report a need for further improvements to their process and dataset. Using Sina Weibo data, 

\cite{guo2015information} proposed a semi-supervised probabilistic graphical model to jointly learn the interactions between user trustworthiness, content reliability, and post credibility for Influenza posts' credibility analysis. Conventional machine learning models (Random Forest and Bayesian Network) were used as baselines for evaluation, and were outperformed by the proposed framework with an accuracy of 71.7\%.

Seeking to curtail the misinformation of COVID-19 related news and support reliable information dissemination, several papers used manual analysis through fact-checkers as well as consensus to verify the veracity and correctness of selected tweets and posts. This is illustrated by a use case in Ethiopia \cite{belay2020towards} analyzing Facebook and Twitter content in both English and Amharic. Similarly, in an Ebola study \cite{sell2020misinformation}, 5\% of Ebola related tweets were found by consensus to be false, while another 5\% contained half-true or misinterpreted information. The authors noted that their findings also indicated "greater than expected politicization of a seemingly neutral international health emergency".

\cite{hossain2020covidlies} approaches misinformation detection from by dividing it into two sub-tasks: retrieval of misconceptions relevant to posts being checked for veracity, and (ii) stance detection to identify whether the posts Agree, Disagree, or express No Stance towards the retrieved misconceptions. This paper provides their dataset of 6761 expert-annotated tweets to evaluate the performance of misinformation detection systems on 86 different pieces of COVID-19 related misinformation. For the first task, the authors report that domain-adaptation and BERTScore (\cite{zhang2019bertscore}), which involves computation over BERT token embeddings of the tweet and misconception, are important for accurate misconception retrieval. As for the second task, findings reported that knowledge about the domain vocabulary helps domain adapted models in predicting the correct stance, as it did for retrieval. 

\textit{\textbf{a.2. Fake news Characterization}}

The work presented in \cite{alsudias2019classifying} used a dataset of 6 million Arabic tweets related to infectious viruses such as MERS and Covid-19 to explore the topics discussed on Twitter in the Arab world during the Covid-19 pandemic, to detect rumors, and to classify the source of tweets into five types of users to determine their veracity. Source classification was realized using logistic regression (LR) to classify tweets into five categories: academic, government, media, health professional and public. Rumor Detection was conducted using a top-down strategy consisting of extracting posts associated with previously identified rumors. Manual annotation was used to create a gold standard dataset and three conventional machine learning models were used for the detection: Logistic regression (LR), Support Vector Classification (SVC) and Naive Bayes (NB). The results of rumor detection reported that he highest accuracy (84.03\%) was achieved by the LR classifier. Interestingly, the source classification results showed that 30\% and 28\% of the rumor tweets' sources were classified as health professional and academic, respectively. The authors explained this result by suggesting that false information often uses language style of academics and health professionals in order to deceive the public. A similar work presented in \cite{alsudias_covid-19_2020} used machine learning classification to identify Covid-19 rumour related tweets with 84\% accuracy, and classified the sources of these rumors using topic modeling based on k-means algorithm.

In an effort to enrich the 'traditional' approach to fake news detection consisting of evaluation text, \cite{zeng2021fake} proposed mining the semantic correlations between the text content and the attached images using a pretrained convolutional neural network (VCG) to learn image representations and use them to enhance textual representations. A combination of these enhanced text representations with a multimodel fusion eigenvector is used to train the fake news detector. This work reported in its results that the model outperformed other approaches on two Twitter and Weibo fake news datasets.

The authors of \cite{shahi2021exploratory} collected a dataset consisting of false or partially false tweets related to Covid-19 from fact-checking websites (e.g., Snopes \footnote{Snopes, Collections archive, 2020. URL: http://www.snopes.com/collections.}), and a random sample of tweets related to COVID-19 from the same period.
To understand how the misinformation around COVID-19 is distinct from the other tweets on this topic, a propagation analysis is performed. Additionally, this study conducts an account categorization in order to gain a better understanding of who is spreading misinformation on Twitter, and analyzes the role of bots in spreading misinformation. The authors report in their findings that false claims propagate faster than partially false claims, and that tweets containing misinformation are more often concerned with discrediting other information on social media. 

The paper presented in \cite{valecha2021fake} proposed to investigate the conditions that lead audiences to accept and disseminate a fake claim as it relates to the Zika virus. The collection of Zika related tweets was conducted in the time period September 2015 - May 2017, while Zika related fake news were collected from various sources. Using neural networks and quantitative content analysis, the authors reported in their findings that Zika tweets that including threat cues and protection cues are positively associated with the likelihood of fake news sharing. 
As an interpretation for that positive association, the authors suggest that fake news reporting higher levels of threat provoke a collective stress reaction, and so, are more likely to be shared in the network. 

A descriptive analysis aiming to understand the ecosystem of information sources shared by Twitter users is presented in \cite{singh2020understanding}. This study conducts a manual categorization of news sources and webpage domains into high-quality health sources, news sources, and low-quality/questionable content providers, based on the URLs contained in tweets. The authors reported in their results that even though the majority of sources have high reliability, a significant proportion do not. This result was used to infer as an indication that the quality of news sources varies considerably with regards to COVID-19 information. 

\cite{sacco2021emergence} used a computational approach along with a Twitter dataset consisting of 200 million interactions captured during the early stage of the pandemic (January-April 2020), to gain insight into the structure of the knowledge communities involved in the creation, filtering and dissemination of COVID-19-related information. 
The paper reported in its results that the Covid-19 infodemic is a highly characteristic community structure, shaped by ideological orientation, typology of fake news, and geographical areas of reference. The URLs appended to messages were labeled according to the political affiliation (left, left-center, neutral, right-center and right) of its media source and the type of source (political, satire, mainstream media, science, conspiracy/junk science, clickbait, fake/hoax). These categories were manually classified by external experts. Additionally, the authors underline a substantial cause for concern in the form of 'troll' accounts, which were found to have the second most prominent role. This finding is interpreted by the authors as a clear sign that "the generation of noise and misinformation is a widespread feature of the digital ecosystem, and a particularly dangerous one from a public health perspective".

\textit{\textbf{a.3. Information Distortion and Conspiracy Theories}}

Using both manual content analysis and topic modeling techniques, \cite{ermakova2020covid} analyzed information distortion types in Twitter cascades. A dataset of 10M tweets in English related to the controversy surrounding Covid-19 medical treatment was collected as part of this work. Manual semantic analysis of tweet content was conducted through an examination of key term distribution and context and of medical terminology verification, allowing for term substitution recognition. Topic modeling was done using Latent Dirichlet Allocation (LDA). The results reported by this work highlighted that distortion and misinformation were caused by oversimplification, distortion of logical links and omission of facts. A shift in the medical topic to political and business disputes was also reported.

In \cite{fan2020social}, a perspective studying risk amplification by information dramatization is offered. The authors argued that the essence of risk spread is that a great deal of information is amplified or weakened by each
recipient and transmitter. This study reported changes on Weibo concerning the focus on the public risk with the development of the epidemic. By exploring and analyzing the relationship between risk communication on media and social amplification of risk on Weibo, the authors conclude that COVID-19 has encountered an appropriate social amplification effect because of dramatic information, controversial topics, as well as social and cultural influences.

In a use case on the misinformation surrounding Covid-19's link to 5G technology, \cite{bahja2020unlink} filtered tweets to 5G from the Twitter dataset presented in \cite{chen2020tracking}, which contains over 50 million COVID-19 related tweets. Latent Dirichlet allocation (LDA) was used to perform a topic modeling task, while a Social Network Analysis approach using centrality and co-occurence analysis of words was used to analyze the different relationships in the tweets' network. The LDA analysis identified several topics from the tweets related to '5G Conspiracy' and '5G Threat' and discussing topics including 5G towers, radiation effects, network and radiation. The authors state that an understanding of the themes and trends from the tweets is crucial for policymakers to counter the misinformation with correct targeted information. This work has a few limitations related to the low number of tweets used in the analysis and the homogeneity of the dataset.

A collection of data from Reddit subreddits and from 4Chan threads related to the pandemic was collected in \cite{shahsavari2020conspiracy} to automatically detect emerging Covid-19 related conspiracy theories.
By estimating narrative networks with an underlying graphical model, the authors analyze the interplay between corpora and track the time-correlation and pervasive flow of information and exploit the latent structure of social media networks and its features to enable the identification of key actors and threat elements in conspiracy theories. Applying the narrative framework discovery pipeline allowed the authors to uncover five central conspiracy theories illustrated by examples such as: Incorporating Covid-19 conspiracy into Q-Anon conspiracy, 5G as the cause of Covid-19, Anti-vax conspiracy and Bill Gates, \#filmyourhospital conspiracy, Pizzagate conspiracy.


\renewcommand{\arraystretch}{2.25}
\begin{small}
\begin{longtable}{p{4.5cm}p{3cm}p{3cm}p{2cm}}
\caption{\textbf{Summary of Methods Used in Papers Addressing Research Question 2: \textit{Social media misinformation during epidemics}}}\\
\label{rq2_tab_fakenews}\\
\rowcolor[rgb]{0.604,0.8,0.8}
\textbf{Method} & \textbf{Epidemic Studied} & \textbf{Social Media Used} & \textbf{References} \\
\rowcolor{white}\multirow{3}{*}~ & ~ & ~ \\[-5ex]
\multirow{3}{*}{\cellcolor[rgb]{0.961,0.961,0.961}}{\textbf{ML Classification}} & {\cellcolor[rgb]{0.961,0.961,0.961}} \multirow{2}{*}{\cellcolor[rgb]{0.961,0.961,0.961}}\textbf{Covid-19} & {\cellcolor[rgb]{0.961,0.961,0.961}} \textbf{Twitter} & {\cellcolor[rgb]{0.961,0.961,0.961}}\cite{abdelminaam2021coaid,madani2021using,alsudias_covid-19_2020}\\ 
\multirow{-2}{*}{\cellcolor[rgb]{0.961,0.961,0.961}} & {\cellcolor[rgb]{0.961,0.961,0.961}} & {\cellcolor[rgb]{0.961,0.961,0.961}} \textbf{Facebook} & {\cellcolor[rgb]{0.961,0.961,0.961}} \cite{maakoul2020towards}\\ 
\rowcolor{white}\multirow{3}{*}{\cellcolor[rgb]{0.961,0.961,0.961}} & ~ & ~ \\[-5ex]
\multirow{-2}{*}{\cellcolor[rgb]{0.961,0.961,0.961}} & {\cellcolor[rgb]{0.961,0.961,0.961}} \textbf{Multiple Epidemics} & \textbf{Twitter} {\cellcolor[rgb]{0.961,0.961,0.961}} & {\cellcolor[rgb]{0.961,0.961,0.961}} \cite{alsudias2019classifying} \\ 
\rowcolor{white}\multirow{3}{*}~ & ~ & ~ \\[-5ex]
\multirow{1}{*}{\cellcolor[rgb]{0.851,0.922,0.922}}{\textbf{DL Classification}}  & {\cellcolor[rgb]{0.851,0.922,0.922}} \multirow{2}{*}{\cellcolor[rgb]{0.851,0.922,0.922}}\textbf{Covid-19} & {\cellcolor[rgb]{0.851,0.922,0.922}} \textbf{Twitter} & {\cellcolor[rgb]{0.851,0.922,0.922}}\cite{abdelminaam2021coaid,zeng2021fake,hossain2020covidlies}\\ 
\multirow{-2}{*}{\cellcolor[rgb]{0.851,0.922,0.922}} & {\cellcolor[rgb]{0.851,0.922,0.922}} & {\cellcolor[rgb]{00.851,0.922,0.922}} \textbf{Weibo} & {\cellcolor[rgb]{0.851,0.922,0.922}} \cite{zeng2021fake} \\ 
\rowcolor{white}\multirow{3}{*}~ & ~ & ~ \\[-5ex]
{\cellcolor[rgb]{0.961,0.961,0.961}}\textbf{Topic Modeling} & {\cellcolor[rgb]{0.961,0.961,0.961}} \textbf{Covid-19}  & {\cellcolor[rgb]{0.961,0.961,0.961}} \textbf{Twitter} & {\cellcolor[rgb]{0.961,0.961,0.961}} \cite{ermakova2020covid,bahja2020unlink}  \\ 
\rowcolor{white}\multirow{3}{*}~ & ~ & ~ \\[-5ex]
{\cellcolor[rgb]{0.851,0.922,0.922}}\textbf{SNA}   & \multirow{2}{*}{\cellcolor[rgb]{0.851,0.922,0.922}}\textbf{Covid-19} & {\cellcolor[rgb]{0.851,0.922,0.922}} \textbf{Twitter} & {\cellcolor[rgb]{0.851,0.922,0.922}} \cite{bahja2020unlink,sacco2021emergence}\\ 
\multirow{-2}{*}{\cellcolor[rgb]{0.851,0.922,0.922}} & {\cellcolor[rgb]{0.851,0.922,0.922}} & {\cellcolor[rgb]{00.851,0.922,0.922}} \textbf{Reddit, 4Chan} & {\cellcolor[rgb]{0.851,0.922,0.922}} \cite{shahsavari2020conspiracy} \\ 
\rowcolor{white}\multirow{3}{*}~ & ~ & ~ \\[-5ex]
{\cellcolor[rgb]{0.961,0.961,0.961}}\textbf{Probabilistic Graph Modeling} & {\cellcolor[rgb]{0.961,0.961,0.961}} \textbf{Influenza}  & {\cellcolor[rgb]{0.961,0.961,0.961}} \textbf{Weibo} & {\cellcolor[rgb]{0.961,0.961,0.961}} \cite{guo2015information}  \\ 
\rowcolor{white}\multirow{3}{*}~ & ~ & ~ \\[-5ex]
\multirow{2}{*}{\cellcolor[rgb]{0.851,0.922,0.922}}\textbf{Manual Content Analysis} & {\cellcolor[rgb]{0.851,0.922,0.922}}\textbf{Covid-19} & {\cellcolor[rgb]{0.851,0.922,0.922}} \textbf{Twitter} & {\cellcolor[rgb]{0.851,0.922,0.922}} \cite{belay2020towards,ermakova2020covid,sacco2021emergence}\\ 
\multirow{-2}{*}{\cellcolor[rgb]{0.851,0.922,0.922}} & {\cellcolor[rgb]{0.851,0.922,0.922}} & {\cellcolor[rgb]{0.851,0.922,0.922}} \textbf{Facebook} & {\cellcolor[rgb]{0.851,0.922,0.922}} \cite{belay2020towards,singh2020understanding} \\
\multirow{-2}{*}{\cellcolor[rgb]{0.851,0.922,0.922}} & {\cellcolor[rgb]{0.851,0.922,0.922}} & {\cellcolor[rgb]{0.851,0.922,0.922}} \textbf{Weibo} & {\cellcolor[rgb]{0.851,0.922,0.922}} \cite{fan2020social} \\
\rowcolor{white}\multirow{3}{*}{\cellcolor[rgb]{0.851,0.922,0.922}} & ~ & ~ \\[-5ex]
\multirow{-2}{*}{\cellcolor[rgb]{0.851,0.922,0.922}} & {\cellcolor[rgb]{0.851,0.922,0.922}} \textbf{Ebola} & {\cellcolor[rgb]{0.851,0.922,0.922}} \textbf{Twitter} & {\cellcolor[rgb]{0.851,0.922,0.922}} \cite{sell2020misinformation} \\ 
\rowcolor{white}\multirow{3}{*}~ & ~ & ~ \\[-5ex]
\multirow{3}{*}{\cellcolor[rgb]{0.961,0.961,0.961}}{\textbf{Quantitative Content Analysis}} & {\cellcolor[rgb]{0.961,0.961,0.961}} \textbf{Covid-19} & {\cellcolor[rgb]{0.961,0.961,0.961}} \textbf{Twitter} & {\cellcolor[rgb]{0.961,0.961,0.961}} \cite{shahi2021exploratory,hossain2020covidlies}\\ 
\multirow{-2}{*}{\cellcolor[rgb]{0.961,0.961,0.961}} & {\cellcolor[rgb]{0.961,0.961,0.961}} & {\cellcolor[rgb]{0.961,0.961,0.961}} \textbf{Weibo} & {\cellcolor[rgb]{0.961,0.961,0.961}} \cite{fan2020social}\\ 
\rowcolor{white}\multirow{3}{*}{\cellcolor[rgb]{0.961,0.961,0.961}} & ~ & ~ \\[-5ex]
\multirow{-2}{*}{\cellcolor[rgb]{0.961,0.961,0.961}} & {\cellcolor[rgb]{0.961,0.961,0.961}} \textbf{Zika} & \textbf{Twitter} {\cellcolor[rgb]{0.961,0.961,0.961}} & {\cellcolor[rgb]{0.961,0.961,0.961}} \cite{valecha2021fake} \\ 
\end{longtable}
\end{small}


\textbf{\textit{Tab. \ref{rq2_tab_fakenews}}} summarizes the methods, epidemics and social media used in studies pertaining to misinformation management and detection.
\subsubsection{RQ3: Can social media be integrated in aspects of public mental health management during epidemics?}

The 2014 Ebola outbreak caused rampant fear behaviors in West Africa \cite{shultz20152014}. The limited health system capacity, coupled with the trauma of witnessing the graphic hemorrhagic manifestations of the virus in the infected, hindered efforts to control the escalating outbreak \cite{shultz20152014}.
Ebola's high infection and mortality rates produced specific mental stressors preventing its survivors from returning to normalcy, as many of them had extreme somatization, obsession-compulsion, depression, anxiety, hostility, phobic anxiety, paranoid ideation, loss of appetite and deterioration of sleep quality \cite{ji2017prevalence}.
The SARS outbreak has created a range of psychiatric conditions including PTSD (Post-Traumatic Stress Disorder), depressive disorders and other anxiety spectrum disorders such as panic, agoraphobia and social phobia \cite{mak2009long}.
The risk for psychological distress does not only threaten survivors and their families, but also frontline workers.
Experience from SARS and H1N1 epidemics underlines that the psychological strain on healthcare professionals could range from reported feelings of extreme vulnerability, uncertainty and threat to life, to somatic and cognitive symptoms of anxiety, and is therefore highly significant \cite{tsamakis2020comment}.

The Covid-19 pandemic "is an individual and collective traumatic event and directly or indirectly has affected every individual in the world" \cite{mukhtar2020psychological}.
Measures taken to curb the spread of the infection, such as lockdown, self-isolation, quarantine and social distancing ,can also be characterized as "a collective traumatic event which poses serious threat to people and have resulted in great loss of lives and property" \cite{mukhtar2020psychological}.
Covid-19 was associated with major stigma and psychological pressure, further aggravating feelings of guilt, shame, regret, sadness, self-pity, anger, internalized emotions, overwhelmed feelings, negative self-talk, unrealistic expectations and perceived sense of failure \cite{mukhtar2020psychological}. Vulnerable populations such as people with pre-existing mental or substance use disorders, people who provide essential services, people infected by the virus are susceptible to the psychological trauma associated with Covid-19 \cite{mukhtar2020psychological}
Populations such as children, seniors, pregnant women, people with disabilities or physical illnesses, abuse victims, people living below the poverty line and other individuals are also high-risk targets for psychological distress due to Covid-19 \cite{mukhtar2020psychological}.

During public health crises such as epidemics and outbreaks, the control strategies put in place to contain the spread of the infection are highly dependent on the transmission method and rate. In the case of MERS, the lower respiratory tract tropism of the disease requires close contact between individuals for the virus to be transmitted from one human to another, thus making healthcare workers and family members at particular risk to acquire secondary MERS infection \cite{widagdo2017mers}. 
Data from the SARS and MERS outbreaks showed that viral shedding and virus excretion peaked approximately 10 to 14 days after the onset of disease \cite{widagdo2017mers}. Consequently, rapid identification and proper quarantine could be highly successful containment measures. The symptomatic or asymptomatic nature of the infection also factors in choosing the adequate control measures to implement. For instance, in the case of SARS, asymptomatic infection was found to be highly unlikely \cite{widagdo2017mers}. In cases where the infection can present as asymptomatic at high percentages, the efficiency of diagnostic screening is limited and so, active surveillance becomes more complicated.
During Covid-19, various containment measures were adopted, including but not limited to: school closures, shut-downs of non-essential businesses, bans on mass gathering, travel restrictions, national border closures, and nationwide curfews \cite{kaimann2021containment}. These measures can worsen mental health state, and contribute to the exacerbation of pre-existing socioeconomic inequalities in mental health \cite{serrano2022impact}.

During epidemics, social media is called upon for various functions ranging from informational support to emotional support and peer support \cite{zhong2021mental}. Given the proliferation in social media use in the last decade, it is not surprising that the use of social media platforms has massively increased during the COVID-19 pandemic \cite{volkmer2021social}, especially with social activities being suspended in many countries. However, this increase in social media use during epidemics can potentially have a negative impact on an already vulnerable mental health.

Two main themes were identified in the selected papers addressing how social media can be integrated in aspects of public mental health management during epidemics, namely: (a) Social media as a tool to gauge mental health toll of epidemics, and (b) Impact of social media consumption during epidemics on mental health. Methods used in each of these themes were classified into groups depending on the aim and results of the study.

\paragraph{\textbf{a. Mental health assessment using social media}}

During the implementation of restrictive measures requiring limiting social contact, social media can become one of the few methods to safely engage with others, rendering it the sole support system of many vulnerable populations. Mental health deterioration can manifest in expressions shared online and be used to gauge the toll epidemics and their containment strategies take on individuals.

The paper presented in \cite{li2020impact} used conventional Machine Learning predictive models and Online Ecological Recognition (OER) to predict the psychological profiles of Weibo users. The authors reported in their findings that anxiety and depression increased while life satisfaction and happiness decreased. 

By sampling and analyzing Weibo posts from 17,865 active users, measures such as word frequency, scores of emotional indicators (e.g., anxiety, depression, indignation, and Oxford happiness) and cognitive indicators (e.g., social risk judgment and life satisfaction) were calculated, and coupled with predictive models based on ecological behavior data from weibo, for the automatic recognition of psychological profiles.
The results reflected that negative emotional
indicators of psychological traits increased in anxiety and depression after the Covid-19 epidemic declaration in China.  

The work presented in \cite{wolohan2020estimating} used LSTM (Long-short Term Memory) neural network text classifier and word embeddings from the fastText library on a Reddit dataset to estimate the population rate of depression in the midst of the COVID-19 pandemic (April 2020). 

The deep LSTM neural network used contains five layers: a fastText embedding layer, three LSTM layers and an output layer. A comparative time-series analysis was performed on three periods, two of which were before the pandemic (January-June 2018, January-June 2019) and one after (January-April 2020). The results reflected a 53\% average increase in depression rate of Reddit users in selected months after the pandemic. 
In an effort to argue the plausibility of their assessment, the authors note that the LSTM used is designed to detect both clinical and sub-clinical depression, and that the user pool used (reddit users) are younger and thus, more prone to depression. The authors also speculate on the association of such depression rates with Covid-related stressors such as stay-at-home orders, potential unemployment and loss of loved ones. 
\renewcommand{\arraystretch}{2.25}
\begin{small}
\begin{longtable}{p{4.5cm}p{3cm}p{3cm}p{2cm}}
\caption{\textbf{Summary of Methods Used in Papers Addressing Research Question 3:} \textit{(a) Social media as a tool to gauge mental health toll of epidemics}}\\
\label{rq3_gaugetoll}\\
\rowcolor[rgb]{0.604,0.8,0.8}
\textbf{Method} & \textbf{Epidemic Studied} & \textbf{Social Media Used} & \textbf{References} \\
\rowcolor{white}\multirow{3}{*}~ & ~ & ~ \\[-5ex]
\multirow{3}{*}{\cellcolor[rgb]{0.961,0.961,0.961}}{\textbf{ML Classification}} & {\cellcolor[rgb]{0.961,0.961,0.961}} \multirow{2}{*}{\cellcolor[rgb]{0.961,0.961,0.961}}\textbf{Covid-19} & {\cellcolor[rgb]{0.961,0.961,0.961}} \textbf{Weibo} & {\cellcolor[rgb]{0.961,0.961,0.961}} \cite{li2020impact}\\ 
\rowcolor{white}\multirow{3}{*}~ & ~ & ~ \\[-5ex]
\multirow{1}{*}{\cellcolor[rgb]{0.851,0.922,0.922}}{\textbf{DL Classification}}  & {\cellcolor[rgb]{0.851,0.922,0.922}} \multirow{2}{*}{\cellcolor[rgb]{0.851,0.922,0.922}}\textbf{Covid-19} & {\cellcolor[rgb]{0.851,0.922,0.922}} \textbf{Reddit} & {\cellcolor[rgb]{0.851,0.922,0.922}} \cite{wolohan2020estimating}\\ 
\rowcolor{white}\multirow{3}{*}~ & ~ & ~ \\[-5ex]
{\cellcolor[rgb]{0.961,0.961,0.961}}\textbf{Topic Modeling} & {\cellcolor[rgb]{0.961,0.961,0.961}} \textbf{Covid-19}  & {\cellcolor[rgb]{0.961,0.961,0.961}} \textbf{Twitter} & {\cellcolor[rgb]{0.961,0.961,0.961}} \cite{viviani2021assessing} \\ 
\end{longtable}
\end{small}

In a vulnerability analysis, the authors of \cite{viviani2021assessing} use Twitter data, topic modeling and expert intervention to evaluate the possible effects of some critical factors related to Covid-19 on the mental well-being of the population. Both top-down and bottom-up approaches are used. The former is based on the definition of target scenarios by experts, while the latter uses a data-driven strategy. The authors report in their findings that psychological vulnerability differs with scenarios. In both approaches, negative psychological vulnerability manifested in negative emotions towards social distancing, hospitalization. 

\textit{\textbf{Tab. \ref{rq3_gaugetoll}}} summarizes the methods, epidemics and social media used in studies pertaining to the use of social media as a tool to gauge the mental health toll of epidemics.

The aforementioned works have a few limitations. Using social media data adds a population/demographic bias to results (\cite{li2020impact,viviani2021assessing}), given that some social media sites are predominantly used by younger people or are more/less popular depending on the country. Moreover, the analysis presented in \cite{li2020impact} is based on a weekly basis, with a relatively large granularity, which has certain influences on reflecting the changing trend of social mentality in a timely manner. The qualitative nature of the results obtained and interpreted by domain experts in \cite{viviani2021assessing} limits the generalisation of the findings and requires more corroborating results, similarly to the findings of \cite{wolohan2020estimating} which need additional data to be strengthened.

\paragraph{\textbf{b. Impact of social media's consumption on mental health}}

Multiple works conducted cross-sectional studies and statistical analysis to gauge the impact of social media use on mental health during epidemics, particularly in the case of Covid-19.

The cross-sectional study presented in \cite{gao2020mental} used an online survey and multivariate logistic regression to analyze the association between social media exposure and mental health problems such as depression and anxiety in Wuhan, China, during the Covid-19 outbreak. The study focused on Sina Weibo in particular. The findings of this study highlight the association between of frequent social media use and higher odds of anxiety, depression, and a combination of both (CDA).  

Similarly, compulsive WeChat use is found to be associated with social media fatigue, emotional stress and social anxiety in the cross-sectional study presented in \cite{pang2021compulsive}. This study was based on the stressor-strain-outcome theoretical paradigm (SSO) and was able to emphasize the mediating role of social media fatigue in the association between Covid-related information overload and psychological outcomes.

The study presented in \cite{zhong2021mental} proposed a conceptual model and used an online survey coupled with regression analysis to to investigate the possible association between social media usage (WeChat in particular) and the mental health toll from the Covid-19 outbreak in Wuhan, China. This study used typical case sampling as an exploratory sampling method to identify the typical cases of Wuhan residents' mental health. Along with social media use, this study also measured emotional support, peer support, health behavior change, depression and secondary trauma. Structural Equation Modeling (SEM) was used to explore the relationship between social media usage and depression and anxiety. The authors reported in their findings an association between social media usage and depression and secondary trauma. Additionally, the authors found that social media usage could significantly predict depression and secondary trauma, indicating that an excessive social media usage contributed to more severe depression and secondary trauma. 
A positive aspect was noted in the study's findings highlighting that social media usage was rewarding to Wuhan's residents through information sharing and emotional and peer support. Indeed, the study explains that excessive use of social media can lead to mental health issues, and that social media breaks has the potential to promote well-being during the pandemic. 

The cross-sectional study presented in \cite{neill2021media} used data collected through an online questionnaire and logistic regression to determine whether COVID-19 related media consumption is associated with changes in mental health outcomes at the beginning of lockdown in the United Kingdom (UK). The data of the survey used the Generalised Anxiety Disorder scale (GAD-7) scale and the Patient Health Questionnaire (PHQ-9), with the baseline data originating from the COVID-19 Psychological Wellbeing Study (\cite{armour2021covid}). The reported findings noted that media usage is statistically significantly associated with anxiety and depression. 

Using an online survey shared via WeChat, an a multivariate logistic regression analysis, the study presented in \cite{ni2020mental} aimed to examine risk factors, including the use of social media, for probable anxiety and depression in the community and among health professionals during the Covid-19 epidemic in Wuhan, China. Depression and anxiety was assessed using by the validated Generalized Anxiety Disorder-2 and Patient Health Questionnaire-2. The study reported finding that close contact with individuals with COVID-19 and spending 2 or more hours daily on COVID-19 related news via social media were associated with probable anxiety and depression in community-based adults.

\renewcommand{\arraystretch}{2.25}
\begin{small}
\begin{longtable}{p{4.5cm}p{3cm}p{3cm}p{2cm}}
\caption{\textbf{Summary of Methods Used in Studies Addressing Research Question 3:} \textit{(b) Impact of Social Media Use on Mental Health during Epidemics}}\\
\label{rq3_mediaconsumption}\\
\rowcolor[rgb]{0.604,0.8,0.8}
\textbf{Method} & \textbf{Epidemic Studied} & \textbf{Social Media Used} & \textbf{References} \\
\rowcolor{white}\multirow{3}{*}~ & ~ & ~ \\[-5ex]
\multirow{3}{*}{\cellcolor[rgb]{0.961,0.961,0.961}}{\textbf{Statistical Analysis}} & {\cellcolor[rgb]{0.961,0.961,0.961}} \multirow{2}{*}{\cellcolor[rgb]{0.961,0.961,0.961}}\textbf{Covid-19} & {\cellcolor[rgb]{0.961,0.961,0.961}} \textbf{WeChat} & {\cellcolor[rgb]{0.961,0.961,0.961}} \cite{zhong2021mental,pang2021compulsive}\\ 
\multirow{-2}{*}{\cellcolor[rgb]{0.961,0.961,0.961}} & {\cellcolor[rgb]{0.961,0.961,0.961}} \multirow{2}{*}{\cellcolor[rgb]{0.961,0.961,0.961}} & {\cellcolor[rgb]{0.961,0.961,0.961}} \textbf{Sina Weibo} & {\cellcolor[rgb]{0.961,0.961,0.961}} \cite{gao2020mental}\\ 
\rowcolor{white}\multirow{3}{*}{\cellcolor[rgb]{0.961,0.961,0.961}} & ~ & ~ \\[-5ex]
\multirow{-2}{*}{\cellcolor[rgb]{0.961,0.961,0.961}} & {\cellcolor[rgb]{0.961,0.961,0.961}} \multirow{2}{*}{\cellcolor[rgb]{0.961,0.961,0.961}}\textbf{Covid-19} & {\cellcolor[rgb]{0.961,0.961,0.961}}\textbf{Social Media} \newline\textit{in general} & {\cellcolor[rgb]{0.961,0.961,0.961}}\cite{neill2021media,ni2020mental,chao2020media,zhao2020social,gao2022social}\\ 

\rowcolor{white}\multirow{3}{*}~ & ~ & ~ \\[-5ex]
\end{longtable}
\end{small}

Similar methods using an online survey and a series of regression analyses were used in \cite{chao2020media}, \cite{zhao2020social} and \cite{gao2022social}, to examine the psychological impact of media use, and found that higher levels of social media use was associated with worse mental health and significantly associated with depression, anxiety and stress. Of particular importance, the results reported in \cite{gao2022social} show that social media exposure is positively associated with anxiety (especially in people with neuroticism) and emotional overeating. 

\textit{\textbf{Tab. \ref{rq3_mediaconsumption}}} summarizes the methods, epidemics, and social media used in studies pertaining to the impact of social media use on mental health during epidemics

The aforementioned studies, although significant, have a few limitations. Firstly, due to their cross-sectional nature, they could not establish a causal relation between media exposure and psychological outcomes, and are reflective of a single point in time for participants, and so, further longitudinal studies are necessary. could not establish a causal relation between media exposure and psychological outcomes.
Secondly, given that these surveys were conducted online, respondent bias is possible (\cite{gao2020mental}). Thirdly, the recruitment of all participants from the same country and from one social media platform can introduce some bias to studies (\cite{pang2021compulsive},\cite{neill2021media}), in addition to gender biases and sample representativeness (\cite{zhong2021mental},\cite{zhao2020social} and \cite{gao2022social}), and recall bias related to self reporting (\cite{chao2020media}). Finally, the results presented do could not exclude the possibility of residual cofounding caused by unmeasured factors.

\section{Discussion}\label{discussion}

This systematic literature review conceptualized three research questions to investigate if, when, and how social media can be used for epidemic management and mitigation, misinformation management, and in the context of public mental health.
As a result of reviewing 129 studies, key themes were identified in relation to each research question, thus providing a systematic analysis of findings for an improved leveraging of social media for successful epidemic management and mitigation, effective curtailment of fake news propagation and negative impact during epidemics, and the effective curtailment of social media's impact on mental health during epidemics.

According to the Behavioral Inhibition System (BIS) theory \cite{berkman2009bas}, people behave in a more reticent and conservative way when they feel threatened by disease. Although epidemics were found to cause negative emotions, many expressions of positive emotions were noted (e.g., \cite{li2020impact}), reflecting group cohesiveness rather than pure personal emotions. It seems that group threats contributed to the manifestation of more beneficial behaviors and social solidarity. Viewing heroic acts, speeches from experts, knowledge of the disease and prevention methods were associated with more positive effects and less expressions of depression \cite{chao2020media}. Media content including useful information for self-protection could be helpful to people during an epidemic outbreak, and may enhance active coping and prevention behaviors which can instill a sense of control \cite{chao2020media}.

The use of social media during epidemics, although linked with manifestations of anxiety and depression, benefited Wuhan residents \cite{zhong2021mental}, and was perceived as an important activity in those lockdown weeks. 
Balancing social media usage in order to obtain ample informational, emotional, and peer support, whilst avoiding the potential mental health toll, is a difficult task for users, especially without the availability of other easy access to other sources of health information \cite{zhong2021mental}.

Although the literature is rich in proposed methods to gauge the role of social media during epidemics and their impact on their management and mitigation, several issues were identified, highlighting research gaps and opportunities for practical implications. 

\subsection{Identified Issues}

One of the major issues identified was the lack of preemptive measures building on the results of previous studies and aiming to implement social media enabled processes in real-time or near real-time. Lessons learned are not efficiently integrated in crisis mitigation measures nor used as building blocks for optimized proactive prevention. A synergy between government health agencies, research communities, and the public would allow for the success of social-media public health initiative in the context of epidemics.

Such collaborative efforts require effective and trust-worthy interactions. This highlights an additional issue related to the relative inefficiency of social media campaigns. Populations need to be targeted for both informative purposes, and for active emotional support. Understanding public opinion is useful to gauge sentiments and reactions, and so it is important to remedy the gap for applications integrating extracted opinions in targeted epidemic management. 

Because of the medical and financial burden of epidemics and outbreaks, mental health concerns are often ignored by both governments and the public. As a result, the manifestation of several mental health related symptoms becomes more prevalent as epidemics progress. In the case of the Ebola outbreak in the year 2014, symptoms of Post-Traumatic Stress Disorder (PTSD) and anxiety-depression were more prevalent even after a year of Ebola response \cite{roy_ebola_2020}. 

When limited resources are geared for epidemic containment, the healthcare system focuses majorly on emergency services. Consequently, individuals suffering from substance abuse and dependency disorders may see deterioration in their mental health as a result \cite{roy2021mental}.  

During community crises, event-related information is often sought in an effort to retain a sense of control in the face of fear and uncertainty and their psychological manifestations. When misleading misinformation is propagated on social media, perceptions of risk are distorted, leading to extreme public panic, stigmatization and marginalization \cite{roy2021mental}. Psychological interventions and psychosocial support would have a direct impact on the improvement of public mental health during epidemics.

\subsection{Directions for Future Research}

As a result of identifying several issues and gaps in the literature related to the research questions of this systematic literature review, we suggest potential paths for future research. 

\begin{itemize}
    \item Given the recognized impact of epidemics on mental health, and the prevalent use of social media platforms during times of crisis, it is necessary to explore the aspects of social media leading to mental health exacerbation during epidemics. Potential factors range from increased consumption levels of social media, social media addiction, emotional fatigue due to overwhelm, and consumption of 'sad' content. Investigating which aspect(s) of social media use are responsible for worsening states of mental health and mental health disorders would allow for a targeted approach to curbing this negative impact during times of crisis. 
    \item In order to manage health-related fake news, it is important to understand what makes citizens prone to engaging in health-related fake news sharing. Specifically, features identifying both an individual's and a group's susceptibility to believe and/or share misinformation need to be determined and categorized. Levels of education, geographical profile, cultural influences, psychological vulnerability and demographic profile are potential features requiring further investigation in their association with fake news dissemination on social media and within communities.
    \item Epidemics are rapidly changing phenomena requiring fast interventions and decision-making. Although post-crisis analysis is imperative for an improved understanding of lessons learned, pro-active epidemic management is vital and would have the most impact on mitigation efforts. Integrating machine learning techniques in this pro-active surveillance could further optimize this process.
    \item Misinformation propagation has a significant impact on the success of interventions given that both the components of exaggerated fear and apathy linked to misinformation can hinder management efforts. However, the investigation of misinformation needs to be extended to include potential links between social media based misinformation and mental health exacerbation. 
\end{itemize}

\subsection{Practical Implications}
This work has several potential practical implications.

\begin{itemize}
    \item \textbf{Implications for governing entities}: (1) develop an efficient misinformation correction strategy to fight incorrect information, rumors and conspiracy theories related to epidemics. (2) develop clear communication channels for knowledge dissemination in order to build trust with the public. (3) develop interventions to limit the impact of epidemics on stress responses (anxiety, depression) due to distorted risk perceptions. (4) Bolster public awareness efforts on sanitary measures and pro-active protection. (5) Ensure the supply of medical staff available to treat patients, as well as psychological support staff to assist patients and their families in navigating the ramifications of infection and of loss of loved ones.
    \item \textbf{Implications for social media platforms}: (1) To take a leadership position in the management of epidemic related fake news by implementing built-in fact checking processes. (2) To assist health agencies and scientific entities in disseminating factual information about the disease, its symptoms, its potential risk and efficient sanitary measures for the public to adopt. 
    \item \textbf{Implications for the public}: Improve community resilience during epidemics using social media groups and assisting in fact dissemination and combatting of misinformation.
\end{itemize}

\section{Conclusion}\label{conclusion}
Given the collective experience of epidemics, responses by communities can often provide insight into the degree of adherence towards preventive measures as well as mitigation protocols. In an effort to control the spread of outbreaks and epidemics, governments, public health institutions and healthcare professionals generally issue guidelines for the public through online portals, news sources and in the past decade, social media. Online "chatter" can indicate the public's response to these guidelines, and their sentiments towards the epidemic itself or specific topics related to it such as vaccinations, treatments, mortality rates, etc. Mitigation efforts require collaborative strategies and public involvement, and so, gaining insight into public opinion and response can prove vital in the success or failure of such efforts.
It is evident that epidemic preparation and mitigation protocols need to be adjusted to deal with the special challenges that accompany the technological revolution taking place, especially in light of the considerable impact of the misinformation infodemic. Additionally, it is vital to have effective ways to exploit the full potential of social media without risking the toll it could potentially take on users' mental health.
The systematic literature review presented in this paper covers several key aspects of the relationship between epidemics, social media and fake news, and various methods used to gauge the impact of such intricate and interconnected elements.

\vspace{6pt} 

\supplementary{PRISMA checklist.}

\authorcontributions{C.A. and M.G. conceived the study. C.A. and I.K. designed the experiments. C.A., I.K. and M.G. carried out the research. C.A. and I.K. prepared the first draft of the manuscript. MG and KB contributed to the experimental design and preparation of the manuscript. All authors were involved in the revision of the draft manuscript and have agreed to the final content.}

\funding{This work was carried out as part of the e-Covid and CovInov projects which are funded by the Centre National de la Recherche Scientifique et Technique (CNRST) and the Rabat-Salé-Kénitra region. }

\conflictsofinterest{The authors declare no conflict of interest.} 






\begin{adjustwidth}{-\extralength}{0cm}

\reftitle{References}
\externalbibliography{yes}
\bibliography{bibliography}

\end{adjustwidth}
\end{document}